\documentclass[twocolumn,twocolappendix]{aastex631}

\usepackage{comment}

\submitjournal{ApJ}

\shorttitle{Vibrationally-excited HC$_{3}$N in the G\,24 disk}
\shortauthors{Taniguchi et al.}
\begin{document}

\title{Vibrationally-excited Lines of HC$_{3}$N Associated with the Molecular Disk\\around the G\,24.78+0.08 A1 Hyper-compact \ion{H}{2} Region}

\correspondingauthor{Kotomi Taniguchi, Kei E. I. Tanaka}
\email{kotomi.taniguchi@nao.ac.jp, kei.tanaka@colorado.edu}

\author[0000-0003-4402-6475]{Kotomi Taniguchi}
\affiliation{National Astronomical Observatory of Japan, National Institutes of Natural Sciences, 2-21-1 Osawa, Mitaka, Tokyo 181-8588, Japan}

\author[0000-0002-6907-0926]{Kei E. I. Tanaka}
\affiliation{Center for Astrophysics and Space Astronomy, University of Colorado Boulder, Boulder, CO 80309, USA}
\affiliation{National Astronomical Observatory of Japan, National Institutes of Natural Sciences, 2-21-1 Osawa, Mitaka, Tokyo 181-8588, Japan}

\author[0000-0001-7511-0034]{Yichen Zhang}
%\affiliation{Star and Planet Formation Laboratory, RIKEN Cluster for Pioneering Research, Hirosawa 2-1, Wako-shi, Saitama 351-0198, Japan}
\affiliation{Department of Astronomy, University of Virginia, Charlottesville, VA 22904, USA}

\author[0000-0003-4040-4934]{Rub\'{e}n Fedriani}
\affiliation{Department of Space, Earth \& Environment, Chalmers University of Technology, 412 93  Gothenburg, Sweden}

\author[0000-0002-3389-9142]{Jonathan C. Tan}
\affiliation{Department of Space, Earth \& Environment, Chalmers University of Technology, 412 93  Gothenburg, Sweden}
\affiliation{Department of Astronomy, University of Virginia, Charlottesville, VA 22904, USA}

\author[0000-0003-0845-128X]{Shigehisa Takakuwa}
\affiliation{Department of Physics and Astronomy, Graduate School of Science and Engineering, Kagoshima University, 1-21-35 Korimoto, Kagoshima, Kagoshima 890-0065, Japan}
\affiliation{Academia Sinica Institute of Astronomy and Astrophysics, 11F of Astro-Math Bldg., 1, Section 4, Roosevelt Road, Taipei 10617, Taiwan}

\author[0000-0001-5431-2294]{Fumitaka Nakamura}
\affiliation{National Astronomical Observatory of Japan, National Institutes of Natural Sciences, 2-21-1 Osawa, Mitaka, Tokyo 181-8588, Japan}
\affiliation{Department of Astronomical Science, School of Physical Science, SOKENDAI (The Graduate University for Advanced Studies), Osawa, Mitaka, Tokyo 181-8588, Japan}
\affiliation{Department of Astronomy, Graduate School of Science, The University of Tokyo, Hongo, Bunkyo, Tokyo 113-0033, Japan}

\author[0000-0003-0769-8627]{Masao Saito}
\affiliation{National Astronomical Observatory of Japan, National Institutes of Natural Sciences, 2-21-1 Osawa, Mitaka, Tokyo 181-8588, Japan}
\affiliation{Department of Astronomical Science, School of Physical Science, SOKENDAI (The Graduate University for Advanced Studies), Osawa, Mitaka, Tokyo 181-8588, Japan}

\author[0000-0001-7031-8039]{Liton Majumdar}
\affiliation{School of Earth and Planetary Sciences, National Institute of Science Education and Research, Jatni 752050, Odisha, India}
\affiliation{Homi Bhabha National Institute, Training School Complex, Anushaktinagar, Mumbai 400094, India}

\author[0000-0002-4649-2536]{Eric Herbst}
\affiliation{Department of Astronomy, University of Virginia, Charlottesville, VA 22904, USA}
\affiliation{Department of Chemistry, University of Virginia, Charlottesville, VA 22904, USA}



\begin{abstract}
We have analyzed Atacama Large Millimeter/submillimeter Array Band 6 data of the hyper-compact \ion{H}{2} region G\,24.78+0.08 A1 (G\,24 HC \ion{H}{2}) and report the detection of vibrationally-excited lines of HC$_{3}$N ($v_{7}=2$, $J=24-23$). %that trace the molecular disk around this massive protostar.
The spatial distribution and kinematics of a vibrationally-excited line of HC$_{3}$N ($v_{7}=2$, $J=24-23$, $l=2e$) are found to be similar to the CH$_{3}$CN vibrationally-excited line ($v_{8}=1$), which indicates that the HC$_{3}$N emission is tracing the disk around the G\,24 HC \ion{H}{2} region previously identified by the CH$_{3}$CN lines.
We derive the $^{13}$CH$_{3}$CN/HC$^{13}$CCN abundance ratios around G\,24 and compare them to the CH$_{3}$CN/HC$_{3}$N abundance ratios in disks around Herbig Ae and T Tauri stars.
The $^{13}$CH$_{3}$CN/HC$^{13}$CCN ratios around G\,24 ($\sim 3.0-3.5$) are higher than the CH$_{3}$CN/HC$_{3}$N ratios in the other disks ($\sim 0.03-0.11$) by more than one order of magnitude.
The higher CH$_{3}$CN/HC$_{3}$N ratios around G\,24 suggest that the thermal desorption of CH$_{3}$CN in the hot dense gas and efficient destruction of HC$_{3}$N in the region irradiated by the strong UV radiation are occurring.
Our results indicate that the vibrationally-excited HC$_{3}$N lines can be used as a disk tracer of massive protostars at the HC \ion{H}{2} region stage, and the combination of these nitrile species will provide information of not only chemistry but also physical conditions of the disk structures.
\end{abstract}


% That way, if you change the order of any elements, LaTeX will
%%% automatically renumber them.

\section{Introduction} \label{sec:intro}

Massive stars ($>8\:M_{\sun}$) play essential roles in the evolution and characterization of galaxies, because they produce and disperse large amounts of energy and heavy elements.
However, the formation processes of massive stars remain uncertain \citep[see, e.g.,][for a review]{2014prpl.conf..149T}.
One formation scenario is the Competitive Accretion model, in which massive stars need to form in clustered environments \citep{2001MNRAS.323..785B, 2010ApJ...709...27W}.
Another is the Turbulent Core Accretion model, a scaled-up version of the formation process of low-mass stars \citep[e.g.,][]{2002Natur.416...59M,2003ApJ...585..850M}, and which can be applied to both clustered and isolated environments.
Theoretical models further investigated stellar feedback from forming massive stars (e.g., disk wind, radiation pressure, photoevaporation, and  stellar winds) in the Core Accretion model \citep[e.g.,][]{2016ApJ...818...52T, 2017ApJ...835...32T}.
In order to reveal the formation process of massive stars and test the theoretical models, high-angular-resolution and high-sensitivity observations are important \citep{2019NatAs...3..517Z, 2019ApJ...886L...4Z}.
In particular, observations of molecular lines are essential, because they provide information about not only gas kinematics but also chemical composition, which is an important tracer of physical conditions and evolutionary stage \citep[e.g.,][]{2012A&ARv..20...56C, 2020ARA&A..58..727J, 2019ApJ...872..154T, 2021ApJ...910..141T}.

Recent high-angular-resolution observations with interferometers such as the Atacama Large Millimeter/submillimeter Array (ALMA) have revealed the presence of disk structures around O-type and B-type massive protostars \citep{2017A&A...602A..59C, 2018A&A...620A..31M, 2019ApJ...873...73Z}.
\citet{2020ApJ...900L...2T} detected the high-temperature components of disks, which are located at very close distances to massive protostars ($\sim 100$ au scale), using several molecular lines such as H$_{2}$O ($v_{2}=1$), NaCl, SiO, and SiS toward the O-type-binary system IRAS\,16547-4247 with ALMA.
However, the detailed physical and chemical properties of such disk structures around massive protostars are still unclear.
The CH$_{3}$CN lines are typically used as a disk tracer \citep{2013A&A...552L..10S,2015ApJ...813L..19J,2016MNRAS.462.4386I}, and sometimes the SiO lines have been used \citep[e.g.,][]{2018A&A...620A..31M,2019ApJ...873...73Z}.

Nitrile species, CH$_{3}$CN and HC$_{3}$N, have been frequently detected in disks around Herbig Ae and T Tauri stars \citep{2018ApJ...857...69B, 2018ApJ...859..131L}.
The Molecules with ALMA at Planet-forming Scale \citep[MAPS;][]{2021ApJS..257....1O} Large Program reveals that these nitrile species trace different layers of these disks; HC$_{3}$N traces upper and warmer layers compared to CH$_{3}$CN \citep{2021ApJS..257....9I}.
In disks around Herbig Ae and T Tauri stars, CH$_{3}$CN is considered to mainly form on dust surfaces followed by the photodesorption or the thermal desorption, while HC$_{3}$N forms in the gas phase \citep{2018ApJ...859..131L, 2019ApJ...886...86L}.
Based on the rotational temperatures of CH$_{3}$CN, the photodesorption is suggested to be more important than the thermal desorption in these disks \citep{2018ApJ...859..131L}.
Hence, it is useful for investigating physical conditions of disks to observe various nitrile species.

In the case of more massive stars, the CH$_{3}$CN lines have been frequently used for searching for disks around massive stars \citep[e.g.,][]{2015ApJ...813L..19J, 2017A&A...602A..59C, 2021arXiv210705683S}, whereas fewer studies about the HC$_{3}$N lines have been conducted.
The vibrationally-excited lines of HC$_{3}$N have been detected in compact hot cores around massive protostars \citep[e.g.,][]{1999A&A...341..882W, 2020ApJ...898...54T}, but it was still unclear whether these lines trace massive protostellar disks due to the insufficient angular resolutions of those observations.
Other higher angular-resolution ALMA data ($\sim 400$ au) show that an HC$_{3}$N vibrationally-excited line ($v_{7}=1e$, $J=38-37$; $E_{\rm {up}}/k=645$ K) traces an accretion disk around the massive protostar in G328.2551-0.5321 \citep{2018A&A...617A..89C}.
Their result suggests that the vibrationally-excited lines of HC$_{3}$N could be a tracer of disks around massive young protostars.
In order to confirm this hypothesis, it is essential to investigate massive protostars in various evolutionary stages (e.g., hyper-compact \ion{H}{2} regions).
In addition, the HC$_{3}$N chemistry in disk structures around massive protostars has not been discussed yet.

In this paper, we present results of two vibrationally-excited HC$_{3}$N ($v_{7}=2$, $J=24-23$, $l=0$ and $l=2e$) lines detected from the G\,24.78+0.08 A1 hyper-compact \ion{H}{2} (hereafter G\,24 HC \ion{H}{2}) region obtained in an ALMA Cycle 6 program.
The bolometric luminosity and distance of this source are $\sim2\times10^{5}$ $L_{\sun}$ and $6.7\pm0.7$ kpc, respectively \citep{2021A&A...650A.142M}.
We adopt a distance of 6.7 kpc to the G\,24 HC \ion{H}{2} region \citep{2021A&A...650A.142M}. 
This source contains a thin shell ionized by an O9.5 star, which was suggested based on studies of the continuum emission and the hydrogen recombination lines \citep{2019A&A...624A.100C}. 
\citet{2021A&A...650A.142M} found that CH$_{3}$CN lines are consistent with a Keplerian rotation around a 20 $M_{\sun}$ star along the axis PA=133\degr (disk major axis), while an ionized jet has been identified along the axis at PA=39\degr\, at the G\,24 HC \ion{H}{2} region.

The structure of this paper is as follows.
We describe the archival data set and reduction procedure in Section \ref{sec:data}.
The resultant moment 0 maps of the molecular lines are presented in Section \ref{sec:res}.
The position-velocity (P-V) diagrams are constructed and kinematics of HC$_{3}$N and CH$_{3}$CN are investigated in Section \ref{sec:pvsub}.
We describe spectral analyses of the $^{13}$CH$_{3}$CN and HC$^{13}$CCN lines and derive the CH$_{3}$CN/HC$_{3}$N abundance ratios around the G\,24 HC \ion{H}{2} region in Sections \ref{sec:ana} and \ref{sec:CH3CN/HC3N}, respectively.
The CH$_{3}$CN/HC$_{3}$N ratios in G\,24 are compared to those in disks around Herbig Ae and T Tauri stars in Section \ref{sec:d2}.
We will discuss the possibility of a binary system in this source in Section \ref{sec:binary}.
In Section \ref{sec:con}, the main conclusions of this paper are summarized.

\section{ALMA Archival Data and Reduction Procedure} \label{sec:data}

We have analyzed the ALMA Band 6 archival data toward the G\,24 HC \ion{H}{2} region\footnote{Project ID; 2018.1.00745.S, PI; Luca Moscadelli}.
The observations were conducted using the 12-m array on July 29 2019 during Cycle 6.
Details of the observations are described in \citet{2021A&A...650A.142M}.

We conducted data reduction and imaging using the Common Astronomy Software Application \citep[CASA;][]{2007ASPC..376..127M} on the pipeline-calibrated visibilities.
We ran the calibration script using CASA version 5.6.1.
The data cubes were created by the tclean task in CASA.
Briggs weighting with a robust parameter of 0.5 was applied.
Pixel size and image size are 0\farcs016 and 3500 $\times$ 3500 pixels, respectively.
The coordinate of the phase center is ($\alpha_{\rm {J}2000}$, $\delta_{\rm {J}2000}$) = (18$^{\rm {h}}$36$^{\rm {m}}$12\fs661, -07\degr12\arcmin10\farcs15).
The maximum baseline length is 8547.6 m.
The field of view (FoV) and maximum recoverable scale (MRS) are approximately 26\arcsec and 0\farcs8, respectively.

Molecular lines and a recombination line presented in this paper are summarized in Table \ref{tab:mol}.
Two vibrationally-excited lines of HC$_{3}$N ($v_{7}=2$, $J=24-23$, $l=0$ and $l=2e$), which are of our main interest, were observed in the same spectral window.
These vibrationally-excited HC$_3$N lines have high upper state energy of $E_{\rm {up}}/k\approx774$--$777$ K, which makes them suitable for tracing inner hot gas near the central star, i.e., the disk.
In addition to the HC$_{3}$N lines, we present the data of the CH$_{3}$CN ($v_{8}=1$, $J_{K,l}=12_{6,1}-11_{6,1}$) line\footnote{The detail studies of this line are summarized in \citet{2021A&A...650A.142M}.} that has a similar upper state energy to the HC$_{3}$N ($v_{7}=2$, $J=24-23$) lines.
We will compare their spatial distributions and P-V diagrams in order to confirm that the HC$_{3}$N lines trace the disk structure around the G\,24 HC \ion{H}{2} region.
We also made data cubes of vibrationally ground-state lines of $^{13}$CH$_{3}$CN ($J_{K}=13_{K}-12_{K}$, $K=0-6$) and HC$^{13}$CCN ($v=0$, $J=26-25$), which are expected to be optically thin, in order to derive their column densities (Section \ref{sec:ana}).
The data cube of the H30$\alpha$ line is presented in Section \ref{sec:binary} to discuss a possibility of the binary system.
The velocity resolution of these data is 0.63\,km\,s$^{-1}$.

Continuum image ($\lambda=1.38$ mm) was created from the broadest spectral window (1.9 GHz bandwidth, center frequency of 217.8 GHz) using the imcontsub task in CASA.
Since some lines have been detected at the HC \ion{H}{2} region position in this spectral window, we determined line-free channels in the tclean image for the imcontsub task.
The polynomial order of 0 was adopted for the continuum estimation.
The rms noise level of the continuum image is 0.1\,mJy\,beam$^{-1}$.

%Table1
\begin{deluxetable}{llcc}
\tabletypesize{\footnotesize}
\tablecaption{Summary of molecular lines and a recombination line presented in this paper \label{tab:mol}}
\tablewidth{0pt}
\tablehead{
\colhead{Species} & \colhead{Transition} & \colhead{Frequency} & \colhead{$E_{\rm {up}}/k$} \\
\colhead{} & \colhead{} &  \colhead{(GHz)} &  \colhead{(K)}
}
%%\decimalcolnumbers
\startdata
HC$_{3}$N ($v_{7}=2$)\tablenotemark{a} & $J=24-23$, $l=0$ & 219.6751141 & 773.5 \\
				        & $J=24-23$, $l=2e$ & 219.7073487 & 776.8 \\
HC$^{13}$CCN ($v=0$)\tablenotemark{a} & $J=26-25$ & 235.5094856 &  152.6 \\				        
CH$_{3}$CN ($v_{8}=1$)\tablenotemark{b} & $J_{K,l}=12_{6,1}-11_{6,1}$ & 221.3118349 & 771.1 \\
$^{13}$CH$_{3}$CN ($v=0$)\tablenotemark{a} & $J_{K}=13_{6}-12_{6}$ & 232.0772032 & 335.5 \\
 								   & $J_{K}=13_{5}-12_{5}$ & 232.1251297 & 256.9 \\
								   & $J_{K}=13_{4}-12_{4}$ & 232.1643692 & 192.5 \\
								   & $J_{K}=13_{3}-12_{3}$ & 232.1949056 & 142.4 \\
								   & $J_{K}=13_{2}-12_{2}$ & 232.2167263 & 106.7 \\
								   & $J_{K}=13_{1}-12_{1}$ & 232.2298223 & 85.2 \\
								   & $J_{K}=13_{0}-12_{0}$ & 232.2341883 & 78.0 \\
CH$_{3}$CN ($v=0$)\tablenotemark{b} & $J_{K}=12_{4}-11_{4}$ & 220.6792869 & 183.1 \\								   
H30$\alpha$						  &          ...                           & 231.900928 & ... \\	    				        
\enddata
\tablenotetext{a}{Rest frequencies were taken from the Cologne Database for Molecular Spectroscopy \citep[CDMS;][]{2005JMoSt.742..215M}.}
\tablenotetext{b}{Rest frequency was taken from the JPL database \citep{1998JQSRT..60..883P}.}
\end{deluxetable}

\section{Results and Discussion} \label{sec:ResAna}

\subsection{Spatial distributions of molecular lines} \label{sec:res} 

\begin{figure*}[!th]
 \begin{center}
  \includegraphics[scale=0.9]{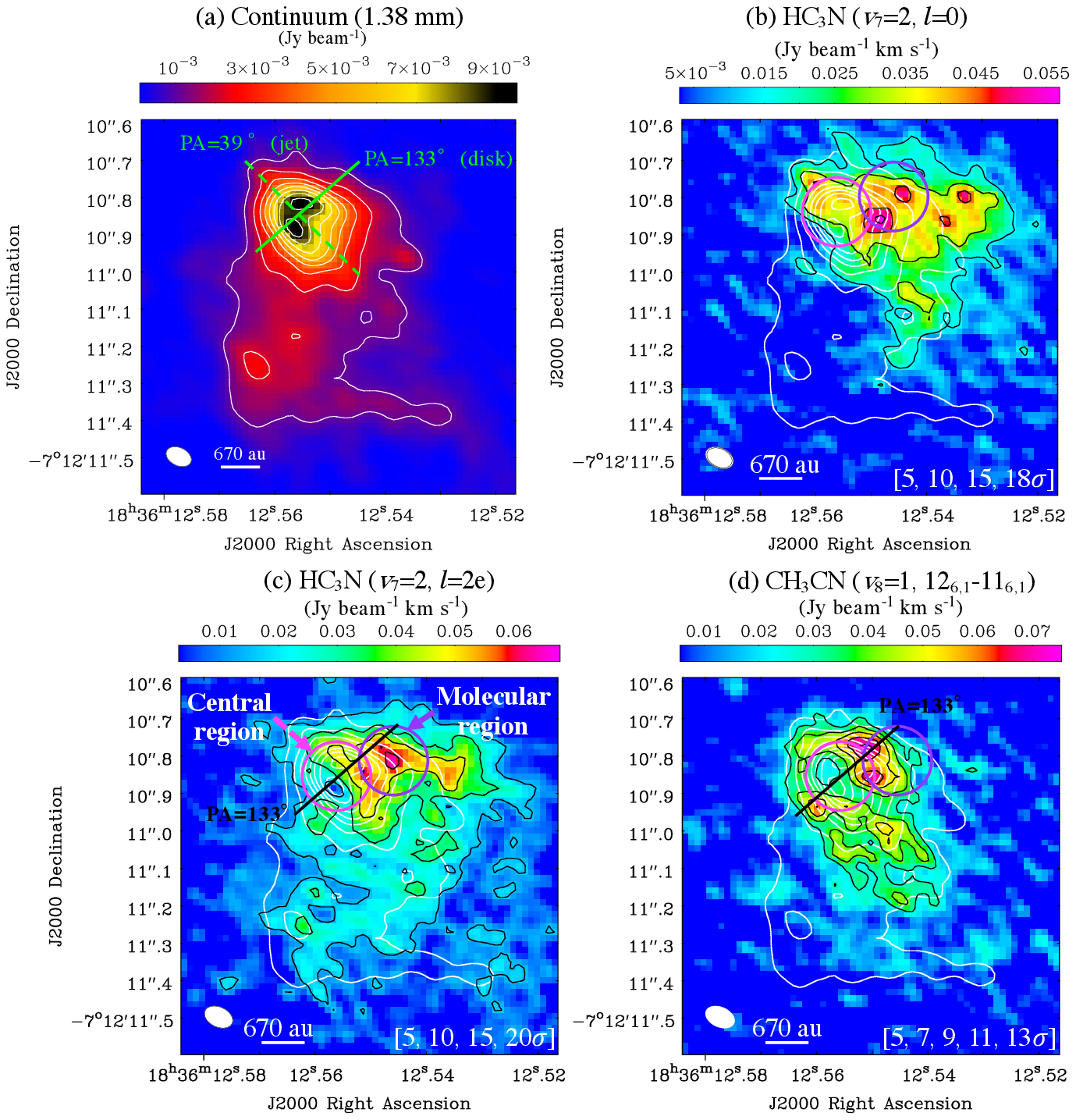}
 \end{center}
\caption{Panel (a) shows the continuum image ($\lambda=1.38$ mm) toward the G\,24 HC \ion{H}{2} region, which was first reported by \citet{2021A&A...650A.142M}. The rms noise level of the image is 0.1 mJy\,beam$^{-1}$. The contour levels are 10, 20, 30, 40, 50, 60, 70, 80 $\sigma$. The filled white ellipse indicates the angular resolution of 0\farcs071 $\times$ 0\farcs048 and PA=63.9\degr. The linear scale is given for 0\farcs1 corresponding to 670 au. The green solid and dashed lines indicates PA=133\degr\, and PA=39\degr, corresponding to the disk major axis and jet directions, respectively \citep{2021A&A...650A.142M}.
Panels (b)--(d) show moment 0 images of HC$_{3}$N ($v_{7}=2$, $J=24-23$, $l=0$), HC$_{3}$N ($v_{7}=2$, $J=24-23$, $l=2e$), and CH$_{3}$CN ($v_{8}=1$, $J_{K,l}=12_{6,1}-11_{6,1}$), respectively, toward the G\,24 HC \ion{H}{2} region. The rms noise levels are 3 mJy\,beam$^{-1}$ km\,s$^{-1}$ for panels (b) and (c), and 5.6 mJy\,beam$^{-1}$ km\,s$^{-1}$ for panel (d), respectively. The black contour levels are indicated at the right bottom in each panel. The white contours indicate the continuum, which is the same one as panel (a). The filled white ellipses indicate the angular resolutions of 0\farcs079 $\times$ 0\farcs053 for panels (b) and (c), and 0\farcs082 $\times$ 0\farcs053 for panel (d), respectively. The linear scale is given for 0\farcs1 corresponding to 670 au. The magenta and purple circles (0\farcs1 radius) indicate the ``Central region'' and ``Molecular region'', respectively. The black lines in panels (c) and (d) indicate the disk major axis (PA=133\degr) for the cutting position in the P-V diagrams (Figure \ref{fig:PV}). \label{fig:mom0}}
\end{figure*}

\begin{figure*}
 \begin{center}
  \includegraphics[width=\textwidth]{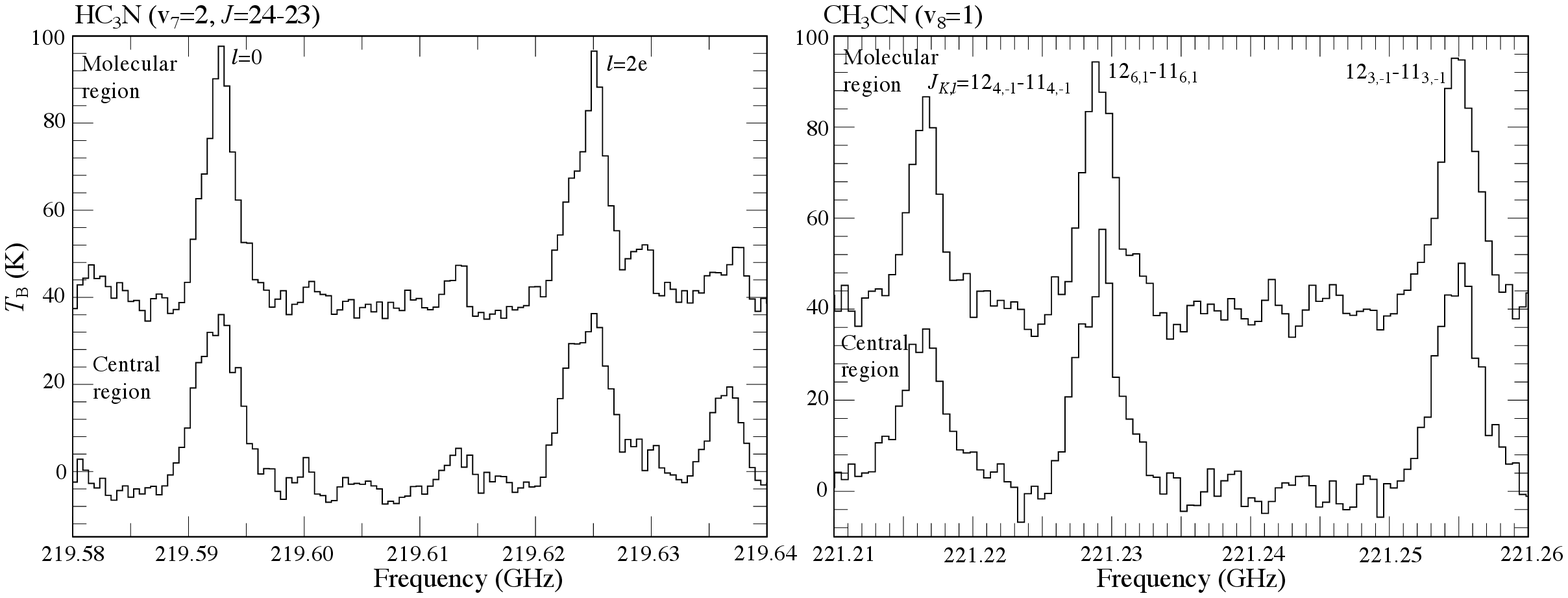}
 \end{center}
\caption{Spectra of the vibrationally-excited lines of HC$_{3}$N (left panel) and CH$_{3}$CN (right panel) at the Central and Molecular regions obtained by averaging over a 0\farcs1 radius. \label{fig:spec_vib}}
\end{figure*}

We show the continuum image ($\lambda=1.38$ mm) toward the G\,24 HC \ion{H}{2} region in panel (a) of Figure \ref{fig:mom0}, which was first reported by \citet{2021A&A...650A.142M}.
The angular resolution is 0\farcs071 $\times$ 0\farcs048, corresponding to $\sim$ 476 au $\times$ 322 au at the source distance of 6.7 kpc.
The beam position angle (PA) is 63.9\degr.
Two peaks are located at the center, and the spatial distribution of the continuum emission is slightly elongated along the north-east to south-west direction (PA = 39\degr).
We will discuss a possible reason for the two peaks in the continuum emission in Section \ref{sec:binary}.
Another continuum core is located $\sim$0\farcs4 (2680 au) south of the strongest continuum core.
This position is consistent with A1M (a molecular emission peak position) named by \citet{2018A&A...616A..66M}, and they found that column densities of several complex organic molecules (COMs) are abundant here.

Panels (b)--(d) of Figure \ref{fig:mom0} show the moment 0 maps of three vibrationally-excited molecular lines; (b) HC$_{3}$N ($v_{7}=2$, $J=24-23$, $l=0$), (c) HC$_{3}$N ($v_{7}=2$, $J=24-23$, $l=2e$), and (d) CH$_{3}$CN ($v_{8}=1$, $J_{K,l}=12_{6,1}-11_{6,1}$); toward the G\,24 HC \ion{H}{2} region.
We select two areas for our analysis, namely ``Central region'' and ``Molecular region'' (panel (c) of Figure \ref{fig:mom0}).
The central position of the Central region corresponds to the middle of the two continuum peaks, and that of the Molecular region is at the peak position of the HC$_{3}$N ($v_{7}=2$, $J=24-23$, $l=2e$) line, respectively.
Their coordinates are ($\alpha_{\rm {J}2000}$, $\delta_{\rm {J}2000}$) = (18$^{\rm {h}}$36$^{\rm {m}}$12\fs556, $-$07\degr12\arcmin10\farcs85), and (18$^{\rm {h}}$36$^{\rm {m}}$12\fs546, $-$07\degr12\arcmin10\farcs80),
respectively.
The other molecular lines also have peaks within the Molecular region.
We selected the two regions to investigate the potential chemical variation induced by different UV fluxes, which is expected to be higher in the Central region.
The Central region is the same position that \citet{2021A&A...650A.142M} analyzed.
Following their work, we adopt the radii of 0\farcs1 for the two regions, which allows us to check the consistency between their and our analyses (see Section \ref{sec:ana}).

Figure \ref{fig:spec_vib} shows the spectra of the HC$_{3}$N and CH$_{3}$CN lines at the Central and Molecular regions.
Other two CH$_{3}$CN lines ($v_{8}=1$, $J_{K,l}=12_{4,-1}-11_{4,-1}$ and $12_{3,-1}-11_{3,-1}$) are also detected near the CH$_{3}$CN ($v_{8}=1$, $J_{K,l}=12_{6,1}-11_{6,1}$) line.
We checked molecular lines in the Splatalogue database\footnote{\url{https://splatalogue.online//advanced1.php}}, and found that no other lines are likely to contaminate the spectra.
In addition, the two vibrationally-excited lines of HC$_{3}$N show similar intensities and spatial distributions as expected.
We thus concluded that line contamination does not occur.

In the moment 0 maps (panels (b) -- (d) of Figure \ref{fig:mom0}),
we can recognize the overall similarities in the spatial distributions of the HC$_{3}$N and CH$_{3}$CN emissions.
Their emission regions have similar extents of $\sim1000{\rm\:au}$,
because they have almost the same upper state energies of $771-777$ K \citep[Table \ref{tab:mol}; see also Figure 6 of][]{2021A&A...650A.142M}.
Also, all three emissions are dimmer in the Central region, probably due to photodissociation by intense UV flux (see also Section \ref{sec:CH3CN/HC3N}).
These similarities suggest that the vibrationally-excited HC$_{3}$N and CH$_{3}$CN lines trace the same hot, dense region surrounding the central ionized region,
i.e., the molecular disk reported by \citet{2021A&A...650A.142M}.
We note, despite these similarities, the HC$_3$N and CH$_3$CN lines also show a difference in their emission morphology.
The CH$_3$CN emission surrounds the Central region, while the two HC$_3$N lines are mainly located in the northwest.
We can explain this spatial difference if the UV flux from the Central region is stronger toward the south direction,
because HC$_3$N would be more efficiently photodissociated than CH$_3$CN (Section \ref{sec:CH3CN/HC3N}).
Unresolved asymmetric innermost structures, e.g., clumpy density structure and a presence of a binary system (Section \ref{sec:binary}), could be at the origin of such an anisotropic UV radiation field.

\subsection{Kinematics of HC$_{3}$N and CH$_{3}$CN} \label{sec:pvsub}

\begin{figure*}[!th]
 \begin{center}
  \includegraphics[width=\textwidth]{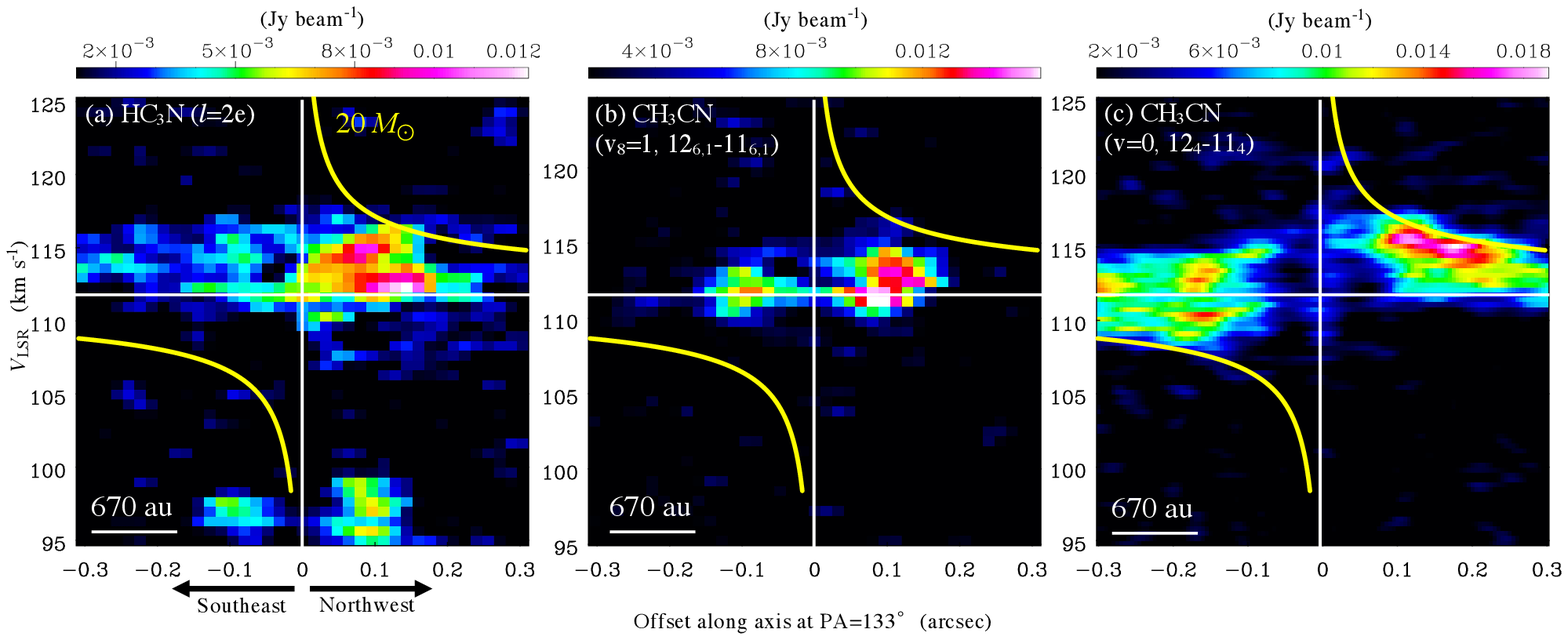}
 \end{center}
\caption{P-V diagrams of (a) HC$_{3}$N ($v_{7}=2$, $J=24-23$, $l=2e$), (b) CH$_{3}$CN ($v_{8}=1$, $J_{K,l}=12_{6,1}-11_{6,1}$), and (c) CH$_{3}$CN ($v=0$, $J_{K}=12_{4}-11_{4}$) along the cut at PA=133\degr (the disk major axis). The horizontal white lines indicate $V_{\rm {LSR}}=112$ km\,s$^{-1}$, which is the systemic velocity of the G\,24 HC \ion{H}{2} region \citep{2021A&A...650A.142M}. The offset center (offset = 0\arcsec) corresponds to the center of the HC \ion{H}{2} region. The positive and negative offsets correspond to northwest and southwest directions, respectively. Yellow lines indicate the Keplerian velocity profile for a central mass of $20\:M_{\odot}$ (the disk inclination is not taken into account). \label{fig:PV}}
\end{figure*}

Having confirmed the similarity of the spatial distributions, we next investigate the kinematical similarity of the vibrationally-excited HC$_3$N and CH$_3$CN lines.
\citet{2021A&A...650A.142M} reported the Keplerian disk of G\,24 using the P-V diagrams of the CH$_{3}$CN lines, including its vibrationally-excited line and $^{13}$C isotopologue line. %$\sim20\:M_{\sun}$.
In order to confirm that the vibrationally-excited HC$_3$N line also traces the same molecular disk,
we constructed P-V diagrams of HC$_{3}$N ($v_{7}=2$, $J=24-23$, $l=2e$) and CH$_{3}$CN ($v_{8}=1$, $J_{K,l}=12_{6,1}-11_{6,1}$) along the disk major axis of PA=133\degr (shown in Figure \ref{fig:mom0}).
We used ``impv'' task in CASA, and averaged the emission over three pixels across the positional cut.
The center position (offset = 0\arcsec) is set at the coordinate of the Central region.
Panels (a) and (b) of Figure \ref{fig:PV} show the P-V diagrams of the vibrationally-excited HC$_{3}$N and CH$_{3}$CN lines, respectively.
Their channel maps in Figures \ref{fig:channelHC3N} and \ref{fig:channelCH3CN} in Appendix \ref{sec:a2}.
For comparison, we also provide the P-V diagram of the vibrational ground-state line of CH$_{3}$CN ($J_{K}=12_{4}-11_{4}$), which clearly shows the Keplerian feature with $20\:M_\odot$
but becomes dimmer in the innermost region (panel (c) of Figure \ref{fig:PV}).

In the P-V diagrams, we could confirm that the HC$_3$N emission has kinematic similarities with the vibrationally-excited and ground-state CH$_3$CN lines, but the resemblance is stronger to the vibrationally-excited line of CH$_{3}$CN.
The brightest features appear in the northwest and redshifted side, and the dimmest features are in the northwest and blueshifted side
in all three panels of Figure \ref{fig:PV}.
On the brightest side, the HC$_3$N emission covers the innermost region of $+0\farcs0 - +0\farcs1$ like the vibrationally-excited CH$_{3}$CN line, which has almost the same upper-state energy (panels (a) and (b)).
Moreover, on the same side, the HC$_3$N emission shows the consistent profile as the Keplerian rotation seen in the ground-state CH$_3$CN emission (panels (a) and (c)).
Hence, we conclude the vibrationally-excited HC$_3$N line associated with
the same Keplerian disk reported by \citet{2021A&A...650A.142M}.
Although the HC$_{3}$N emission shows some resemblance to that of CH$_{3}$CN in the P-V diagrams (Figure \ref{fig:PV}), the degree of this resemblance is still debatable.
Future higher-angular resolution observations including other transitions are necessary for further discussions.

The other similarity is that these HC$_3$N and CH$_3$CN lines show emission features
on the southwest and redshifted side of the P-V diagrams \citep[see also Figure 4 of ][]{2021A&A...650A.142M}.
The distribution of the HC$_3$N emission lines is intermediate between those of the vibrationally-excited and ground-state CH$_3$CN lines.
This again supports that the HC$_3$N line traces the same gas as CH$_3$CN.
We note that such a kinematical feature cannot be reproduced by a pure Keplerian disk \citep{2014Natur.507...78S, 2019ApJ...873...73Z}.
Contamination from an infalling, rotating envelope is able to create this feature, even though we used the lines with the high upper-state energies,
which can be excited only in hot dense gas \citep[cf., $E_{\rm {up}}/k=777$ K and the critical density of $\gtrsim 4 \times 10^{8}{\rm\:cm}^{-3}$ for the HC$_3$N line;][]{1999A&A...341..882W}.
One difference between the HC$_3$N and ground-state CH$_3$CN lines appears on the southeast and blueshifted sides in their P-V diagrams (panels (a) and (c)).
The HC$_3$N emission is darker on the southwest side, making it difficult to trace that side of the Keplerian feature (see also Figure \ref{fig:mom0}).
An anisotropic UV radiation field could create this HC$_3$N depletion, which we also mentioned in Section \ref{sec:res}.
This would imply that fast chemical processes change the chemical composition of the molecular disk with a shorter timescale than the rotational timescale for the disk \citep[e.g.,][]{2017ApJ...843L...3C}.

\subsection{The abundances of HC$_{3}$N and CH$_{3}$CN} \label{sec:13C}

As mentioned in Section \ref{sec:intro}, HC$_{3}$N and CH$_{3}$CN trace different layers of disks around Herbig Ae and T Tauri stars \citep{2021ApJS..257....9I}, suggesting that these species trace different physical conditions in disk structures.
In order to investigate the physical and chemical conditions of the molecular disk around the G\,24 HC \ion{H}{2} region, we will evaluate the CH$_{3}$CN/HC$_{3}$N abundance ratios.

\subsubsection{Spectral line analyses of $^{13}$CH$_{3}$CN and HC$^{13}$CCN} \label{sec:ana}

\begin{figure*}[!th]
 \begin{center}
  \includegraphics[scale=0.73]{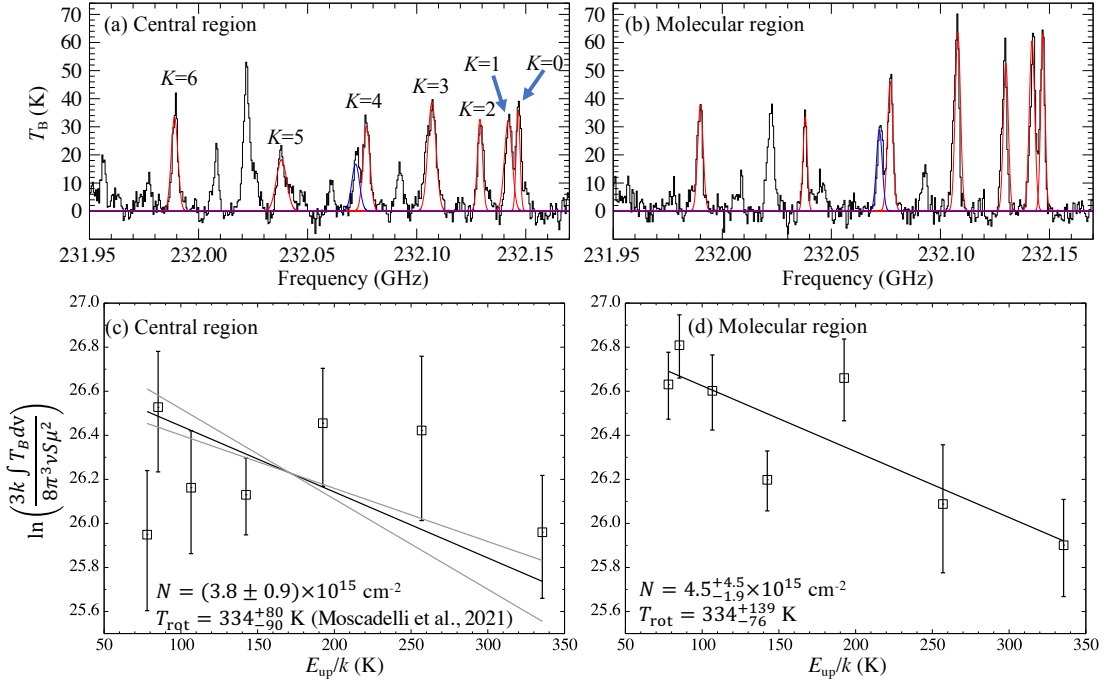}
 \end{center}
\caption{Spectra of $^{13}$CH$_{3}$CN obtained by averaging over a 0\farcs1 radius at the center of the Central and Molecular regions in panels (a) and (b), respectively. Red curves show the Gaussian fitting results for the $^{13}$CH$_{3}$CN lines, and blue curves indicate the fit results for a line from another molecule (we could not clearly identify a molecule and its transition because of a possibility of line contamination). Panels (c) and (d) show results of the rotational diagram fit at the Central and Molecular regions, respectively. The error bars for each data point are derived from the standard deviation of the line fitting with a Gaussian profile summarized in Table \ref{tab:gaussfit}. In panel (c), the black line shows the fitting result to our data using $T_{\rm {rot}}=334$ K, and the gray lines indicate the fitting results to our data using $T_{\rm {rot}}$=($334+80$) K and $T_{\rm {rot}}$=($334-90$) K. \label{fig:CH3CN}}
\end{figure*}

%Table2
\begin{deluxetable}{cccccc}
\tablecaption{Gaussian fitting results of the $^{13}$CH$_{3}$CN and HC$^{13}$CCN \label{tab:gaussfit}}
\tablewidth{0pt}
\tablehead{
\colhead{} & \multicolumn{2}{c}{Central region} & \colhead{} & \multicolumn{2}{c}{Molecular region} \\
\cline{2-3} \cline{5-6}
\colhead{Line} & \colhead{$T_{\rm {B}}$ (K)} & \colhead{$\Delta v$ (km\,s$^{-1}$)} & \colhead{} & \colhead{$T_{\rm {B}}$ (K)} & \colhead{$\Delta v$ (km\,s$^{-1}$)}
}
%%\decimalcolnumbers
\startdata
\multicolumn{6}{c}{{\bf {$^{13}$CH$_{3}$CN ($J_{K}=13_{K}-12_{K}$)}}} \\
$K=0$ & $36.0 \pm 4.4$ & $2.7 \pm 0.5$ & & $63.1 \pm 3.7$ & $3.0 \pm 0.3$ \\
$K=1$ & $32.1 \pm 3.1$ & $5.4 \pm 0.9$ & & $60.6 \pm 3.4$ & $3.8 \pm 0.3$ \\
$K=2$ & $32.6 \pm 3.8$ & $3.6 \pm 0.6$ & & $52.6 \pm 3.7$ & $3.5 \pm 0.3$ \\
$K=3$ & $38.7 \pm 3.0$ & $5.7 \pm 0.5$ & & $63.5 \pm 3.7$ & $3.7 \pm 0.3$ \\
$K=4$ & $30.2 \pm 3.3$ & $4.8 \pm 0.8$ & & $46.1 \pm 3.4$ & $3.9 \pm 0.4$ \\
$K=5$ & $18.3 \pm 2.7$ & $7.3 \pm 1.6$ & & $33.6 \pm 3.9$ & $2.8 \pm 0.5$ \\
$K=6$ & $33.5 \pm 3.4$ & $4.6 \pm 0.8$ & & $37.6 \pm 3.4$ & $3.9 \pm 0.5$ \\
\multicolumn{6}{c}{{\bf {HC$^{13}$CCN ($J=26-25$)}}} \\
$J=26-25$ & $33.4 \pm 1.7$ & $6.0 \pm 0.4$ & & $58.1 \pm 3.0$ & $4.6 \pm 0.3$
\enddata
\tablecomments{Errors indicate the standard deviation.}
\end{deluxetable}

We derive column densities of HC$^{13}$CCN and $^{13}$CH$_{3}$CN using their vibrational ground-state lines.
Figure \ref{fig:13C_mom0} in Appendix \ref{sec:a1} shows the moment 0 maps of the vibrational ground-state lines of their $^{13}$C isotopologues.
These lines are usually optically thin, and we can derive their column densities more accurately than the case with the main species.

We obtained the spectra of $^{13}$CH$_{3}$CN and HC$^{13}$CCN at the Central and Molecular regions over the 0\farcs1 radius.
As described in Section \ref{sec:res}, the Central region is the same region where \citet{2021A&A...650A.142M} analyzed.
The derived column densities and rotational temperatures in this section are the beam-averaged values, as \citet{2021A&A...650A.142M} derived.

Panels (a) and (b) of Figure \ref{fig:CH3CN} show the spectra of $^{13}$CH$_{3}$CN ($J=13-12$) at the Central and Molecular regions, respectively.
We fitted the spectra with a Gaussian profile, and summarized the obtained parameters in Table \ref{tab:gaussfit}.
We analyzed the $^{13}$CH$_{3}$CN spectra with the rotational diagram method using the following formula \citep{1999ApJ...517..209G};
\begin{equation} \label{rd}
{\rm {ln}} \frac{3k \int T_{\mathrm {B}}dv}{8\pi ^3 \nu S \mu ^2} = {\rm {ln}} \frac{N}{Q(T_{\rm {rot}})} - \frac{E_{\rm {up}}}{kT_{\rm {rot}}},
\end{equation}
where $k$ is the Boltzmann constant, $S$ is the line strength, $\mu$ is the permanent electric dipole moment, $N$ is the column density, $T_{\rm {rot}}$ is the rotational temperature, $E_{\rm {up}}$ is the upper-state energy, and $Q(T_{\rm {rot}})$ is the partition function.
The permanent electric dipole moment of $^{13}$CH$_{3}$CN is 3.92197 Debye \citep{1995JCP....92.1984G}.

Panels (c) and (d) of Figure \ref{fig:CH3CN} show results of the rotational diagram analysis at the Central and Molecular regions, respectively.
At the Central region, we could not obtain the rotational temperature by the fitting, partly because some lines do not show a Gaussian profile, and then we applied an rotational temperature of $334^{+80}_{-90}$ K derived by \citet{2021A&A...650A.142M}.
Their results are applicable, since \citet{2021A&A...650A.142M} analyzed the $^{13}$CH$_{3}$CN spectra at the HC \ion{H}{2} region within a 0\farcs1 radius, which is the same region as we have analyzed.
The column density of $^{13}$CH$_{3}$CN at the HC \ion{H}{2} region is derived to be ($3.8 \pm 0.9$)$\times 10^{15}$ cm$^{-2}$. 
This column density is consistent with the previous result within the uncertainties \citep[$5.01^{+1.2}_{-1.5}\times10^{15}$ cm$^{-2}$;][]{2021A&A...650A.142M}.
The rotational temperature and column density at the Molecular region are derived to be $334^{+139}_{-76}$ K and $4.5^{+4.5}_{-1.9}\times10^{15}$ cm$^{-2}$, respectively, by the rotational diagram fitting for our data.

The derived rotational temperatures at the two positions are much higher than the sublimation temperature of CH$_{3}$CN ($\approx 95$ K\footnote{We took the sublimation temperature from a hot-core model by \citet{2019ApJ...881...57T}.}), which is different from results in the disks around the Herbig Ae and T Tauri stars \citep{2018ApJ...857...69B, 2018ApJ...859..131L, 2021ApJS..257....9I}.
Such high rotational temperatures of $^{13}$CH$_{3}$CN indicate that CH$_{3}$CN, including its isotopologues, thermally desorbs from dust grains rather than photodesorption by the UV radiation around the G\,24 HC \ion{H}{2} region.

\begin{figure}[!th]
 \begin{center}
  \includegraphics[bb =0 30 377 504, scale=0.5]{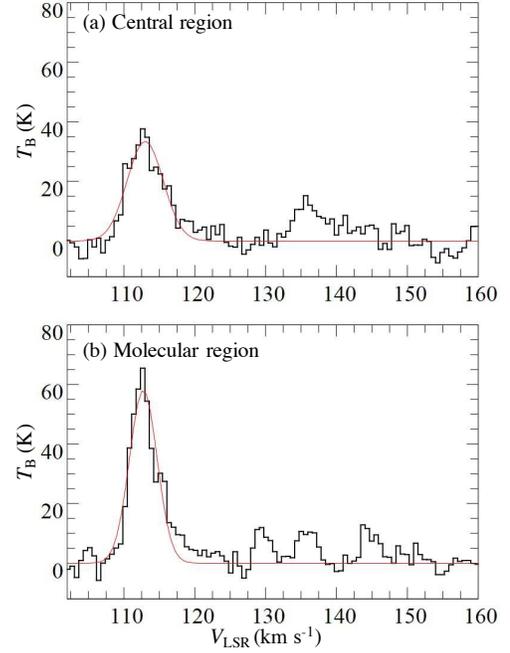}
 \end{center}
\caption{Spectra of HC$^{13}$CCN ($J=26-25$) obtained by averaging over a 0\farcs1 radius at the center of the Central and Molecular regions in panels (a) and (b), respectively. Red curves show the results of the gaussian fit. \label{fig:HC13CCNspec}}
\end{figure}

%Table3
\begin{deluxetable*}{ccccccc}
\tablecaption{Column density and optical depth of HC$^{13}$CCN \label{tab:columnHC13CCN}}
\tablewidth{0pt}
\tablehead{
\multicolumn{3}{c}{Central region} & \colhead{} & \multicolumn{3}{c}{Molecular region}\\
\cline{1-3} \cline{5-7}
\colhead{$T_{\rm {ex}}$ (K)\tablenotemark{a}} & \colhead{$N$ (cm$^{-2}$)} & \colhead{$\tau$} & \colhead{} & \colhead{$T_{\rm {ex}}$ (K)\tablenotemark{a}} & \colhead{$N$ (cm$^{-2}$)} & \colhead{$\tau$}
}
%%\decimalcolnumbers
\startdata
334 & ($1.08 \pm 0.08$)$\times 10^{15}$ & 0.107 & & 334 & ($1.48 \pm 0.12$)$\times 10^{15}$ & 0.195 \\
244 & ($9.4 \pm 0.7$)$\times 10^{14}$ & 0.151 & & 258 & ($1.36 \pm 0.11$)$\times 10^{15}$ & 0.262 \\
414 & ($1.20 \pm 0.09$)$\times 10^{15}$ & 0.086 & & 473 & ($1.78 \pm 0.14$)$\times 10^{15}$ & 0.133
\enddata
\tablecomments{Errors indicate the standard deviation.}
\tablenotetext{a}{Excitation temperatures ($T_{\rm {ex}}$) are taken from the results of $^{13}$CH$_{3}$CN.}
\end{deluxetable*}

Figure \ref{fig:HC13CCNspec} shows spectra of the HC$^{13}$CCN ($J=26-25$) line at the Central and Molecular regions, obtained by the same method as that for $^{13}$CH$_{3}$CN.
We fitted the spectra with a Gaussian profile, and the fitting results are shown as red curves in Figure \ref{fig:HC13CCNspec}.
Table \ref{tab:gaussfit} summarizes the line parameters obtained by the fitting.
The $V_{\rm {LSR}}$ values of this line are $113.0 \pm 0.15$ km\,s$^{-1}$ and $112.7 \pm 0.12$ km\,s$^{-1}$ at the Central and Molecular regions, respectively.

\begin{figure}[!th]
 \begin{center}
  \includegraphics[bb =20 30 418 316, scale=0.5]{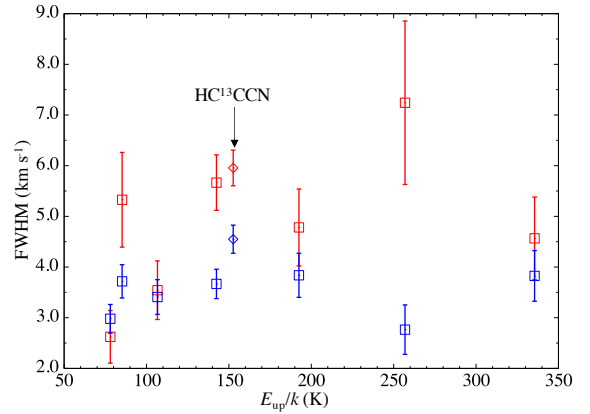}
 \end{center}
\caption{Plot of the upper state energy ($E_{\rm {up}}$) vs. the line width (FWHM) of the $^{13}$CH$_{3}$CN and HC$^{13}$CCN lines. Red and blue plots represent data points of the Central and Molecular regions, respectively. The square and diamond signs indicate the $^{13}$CH$_{3}$CN and HC$^{13}$CCN data, respectively. \label{fig:FWHM}}
\end{figure}

Since we have only one rotational transition line of HC$^{13}$CCN, we need a reasonable assumption of its excitation temperature to derive its column density. 
If the emission region of $^{13}$CH$_{3}$CN coincides with that of HC$^{13}$CCN and the LTE condition is achieved, we can use the excitation temperature of $^{13}$CH$_{3}$CN as the reference value.
Therefore, we compare their line widths of $^{13}$CH$_{3}$CN and HC$^{13}$CCN to examine their spatial coincidence.

Figure \ref{fig:FWHM} shows relationships between the line width (FWHM) and the upper state energy of the $^{13}$CH$_{3}$CN and HC$^{13}$CCN lines.
We derive the line width (FWHM) using 
\begin{equation}
{\rm {FWHM}} = \sqrt{(\Delta v_{\rm {obs}})^2 - (\Delta v_{\rm {inst}})^{2}},
\end{equation}
where $\Delta v_{\rm {obs}}$ and $\Delta v_{\rm {inst}}$ are the observed line widths and the instrumental velocity width (0.63 km\,s$^{-1}$), respectively.
There are no clear dependencies of the line width on the upper state energy.
The value of the $K=5$ line of $^{13}$CH$_{3}$CN and its error toward the Central region are slightly larger than those of the other lines, which may be due to the contamination of another line (panel (a) of Figure \ref{fig:CH3CN}).
The line widths at the Central region are generally larger than those at the Molecular region, and the line widths of the HC$^{13}$CCN line are comparable to those of $^{13}$CH$_{3}$CN.
Thus, we hereafter assume the excitation temperature of HC$^{13}$CCN is the same as the $^{13}$CH$_{3}$CN's rotational temperature.

We derived the column densities of HC$^{13}$CCN at the two positions assuming the LTE condition \citep{1999ApJ...517..209G}.
We used the following formulae;
\begin{equation} \label{tau}
\tau = - {\mathrm {ln}} \left[1- \frac{T_{\rm {B}} }{J(T_{\rm {ex}}) - J(T_{\rm {bg}})} \right]
\end{equation}
where
\begin{equation} \label{tem}
J(T) = \frac{h\nu}{k}\Bigl\{\exp\Bigl(\frac{h\nu}{kT}\Bigr) -1\Bigr\} ^{-1},
\end{equation}  
and
\begin{eqnarray} \label{col}
N = \tau \frac{3h\Delta v}{8\pi ^3}\sqrt{\frac{\pi}{4\mathrm {ln}2}}Q(T_{\rm {ex}})\frac{1}{\mu ^2}\frac{1}{J_{\rm {lower}}+1} \nonumber \\ \exp\Bigl(\frac{E_{\rm {lower}}}{kT_{\rm {ex}}}\Bigr) \Bigl\{1-\exp\Bigl(-\frac{h\nu }{kT_{\rm {ex}}}\Bigr)\Bigr\} ^{-1}.
\end{eqnarray} 
In Equation (\ref{tau}), $\tau$ denotes the optical depth, and $T_{\rm {B}}$ the peak intensities summarized in Table \ref{tab:gaussfit}.
$T_{\rm{ex}}$ and $T_{\rm {bg}}$ are the excitation temperature and the cosmic microwave background temperature (2.73 K).
We calculated three cases of excitation temperatures ($T_{\rm {ex}}$, $T_{\rm {ex}}\pm T_{\rm {ex}}$(error)) for each position.
$J$($T$) in Equation (\ref{tem}) is the effective temperature equivalent to that in the Rayleigh-Jeans law.
In Equation (\ref{col}), {\it N} is the column density,  $\Delta v$ is the line width, $Q$($T_{\rm {ex}}$) is the partition function at $T_{\rm {ex}}$, $\mu$ is the permanent electric dipole moment, and $E_{\rm {lower}}$ is the energy of the lower rotational energy level. 
The electric dipole moment of HC$^{13}$CCN is 3.732 Debye \citep[CDMS database;][]{2005JMoSt.742..215M}.

Table \ref{tab:columnHC13CCN} summarizes the derived column densities of HC$^{13}$CCN and the peak optical depth ($\tau$) at each position.
We confirmed that the HC$^{13}$CCN line is optically thin.
The derived column densities at the Molecular region are slightly higher than those at the Central region, as in the case of $^{13}$CH$_{3}$CN (see Figure \ref{fig:CH3CN}).

\subsubsection{The CH$_{3}$CN/HC$_{3}$N abundance ratios in G\,24} \label{sec:CH3CN/HC3N}

\begin{figure*}[!th]
 \begin{center}
  \includegraphics[bb = 0 20 431 235, scale=0.85]{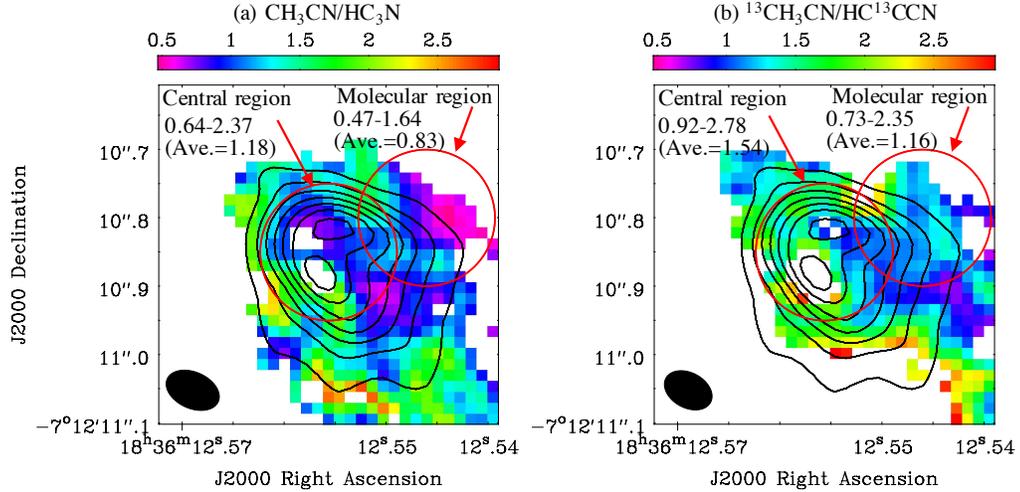}
 \end{center}
\caption{Maps of the moment 0 ratios between (a) CH$_{3}$CN ($v_{8}=1$, $J_{K,l}=12_{6,1}-11_{6,1}$) and HC$_{3}$N ($v_{7}=2$, $J=24-23$, $l=2e$) (panels (d) and (c) of Figure \ref{fig:mom0}, respectively) and (b) $^{13}$CH$_{3}$CN ($J_{K}=13_{3}-12_{3}$) and HC$^{13}$CCN ($J=26-25$) (panels (c) and (a) of Figure \ref{fig:13C_mom0}, respectively) toward G\,24. The red circles represent the Central and Molecular regions. The indicated values are minimum, maximum and average values in each region. The average values are calculated based on the total fluxes in each moment 0 map. The black contours indicate the continuum emission from $20\sigma$ to $80\sigma$. \label{fig:ratiomap}}
\end{figure*}

The $N$($^{13}$CH$_{3}$CN)/$N$(HC$^{13}$CCN) ratios are derived to be $3.47 \pm 0.46$ and $3.10_{-1.62}^{+3.68}$ at the Central and Molecular regions, respectively, using the results in Section \ref{sec:ana}.
If there are no effects of the $^{13}$C isotopic fractionation in each molecule \citep[e.g.,][]{2016ApJ...830..106T, 2017ApJ...846...46T} and the selective photodissociation of $^{13}$CH$_{3}$CN and HC$^{13}$CCN, the $N$($^{13}$CH$_{3}$CN)/$N$(HC$^{13}$CCN) ratio reflects the CH$_{3}$CN/HC$_{3}$N abundance ratio.
In the case of G\,24 with the strong UV radiation field, it is likely that the effect of the $^{13}$C isotopic fractionation of HC$_{3}$N is negligible \citep{2019ApJ...881...57T,2021ApJ...908..100T}.

In order to confirm the CH$_{3}$CN/HC$_{3}$N abundance ratios around G\,24, we investigated the CH$_{3}$CN/HC$_{3}$N ratios using the moment 0 maps of their vibrationally-excited lines.
These lines have similar upper state energies ($\sim 770$ K, Table \ref{tab:mol}).
In addition, the estimated gas density traced by the CH$_{3}$CN ($v_{8}=1$) lines is around $10^{9}$ cm$^{-3}$ \citep{2021A&A...650A.142M}, and the critical density of the HC$_{3}$N ($v_{7}=2$) lines is $\geq 4 \times 10^{8}$ cm$^{-3}$ \citep{1999A&A...341..882W}.
Hence, these vibrationally-excited lines of CH$_{3}$CN and HC$_{3}$N likely trace similar regions around G\,24, and their excitation conditions are expected to be equivalent.
If these lines are optically thin, this integrated-intensity ratio would be comparable to their abundance ratio.
Panel (a) of Figure \ref{fig:ratiomap} shows the map of the ratio between the CH$_{3}$CN moment 0 map (panel (d) of Figure \ref{fig:mom0}) and that of HC$_{3}$N (panel (c) of Figure \ref{fig:mom0}) toward G\,24.
We included only pixels with above $3\sigma$ detection of both species.
We also show the same map but between the $^{13}$CH$_{3}$CN ($J_{K}=13_{3}-12_{3}$) and HC$^{13}$CCN ($J=26-25$) lines (panels (c) and (a) of Figure \ref{fig:13C_mom0}, respectively), in order to cross check the results.
We used the $^{13}$CH$_{3}$CN ($J_{K}=13_{3}-12_{3}$) line, because it has the closest upper-state energy to the HC$^{13}$CCN ($J=26-25$) line (see Table \ref{tab:mol}).

The CH$_{3}$CN/HC$_{3}$N ratios including their $^{13}$C isotopologues cannot be derived near the continuum peak due to the below $3\sigma$ detection of HC$_{3}$N, which could indicate the efficient HC$_{3}$N destruction at the Central region. 
In panel (a), the CH$_{3}$CN/HC$_{3}$N ratio shows lower values in the Molecular region ($\approx0.5-1.6$, the average value is 0.83) compared to those in the Central region ($\approx0.6-2.4$, the average value is 1.18).
This marginal difference in the CH$_{3}$CN/HC$_{3}$N ratio between the two regions suggests that the UV photodissociation and/or reactions with ions increase the CH$_{3}$CN/HC$_{3}$N abundance ratio in the region irradiated by the strong UV radiation.
This can be caused by the following two reasons;
\begin{enumerate}
    \item HC$_{3}$N could be more efficiently destroyed at the Central region, because the UV photodissociation rate of HC$_{3}$N is higher than that of CH$_{3}$CN by a factor of $\sim2.4$ \citep[Table 2 in][]{2019ApJ...886...86L}.
    \item HC$_{3}$N is destroyed by reactions with ions (H$^{+}$, H$_{3}^{+}$, HCO$^{+}$), which should be abundant in ionized regions \citep{2019ApJ...881...57T}.
\end{enumerate}

The $^{13}$CH$_{3}$CN/HC$^{13}$CCN ratios are derived to be $\approx0.9-2.8$ (the average value is 1.54) and $\approx0.7-2.4$ (the average value is 1.16) at the Central and Molecular regions, respectively (panel (b) of Figure \ref{fig:ratiomap}).
The $^{13}$CH$_{3}$CN/HC$^{13}$CCN ratio at the Molecular region is lower than that at the Central region, as seen in panel (a), and this tendency is also consistent with the $N$($^{13}$CH$_{3}$CN)/$N$(HC$^{13}$CCN) ratios.

We adopt the $N$($^{13}$CH$_{3}$CN)/$N$(HC$^{13}$CCN) column density ratio as representative values in G\,24 in the following section, because the CH$_{3}$CN/HC$_{3}$N ratio based on the moment 0 maps may be affected by the effect of the optically thickness. 

\subsubsection{Comparisons of the CH$_{3}$CN/HC$_{3}$N abundance ratios among the high- and lower-mass protostellar disks} \label{sec:d2}

\begin{figure*}[!th]
 \begin{center}
 \includegraphics[bb = 20 30 542 439, scale=0.55]{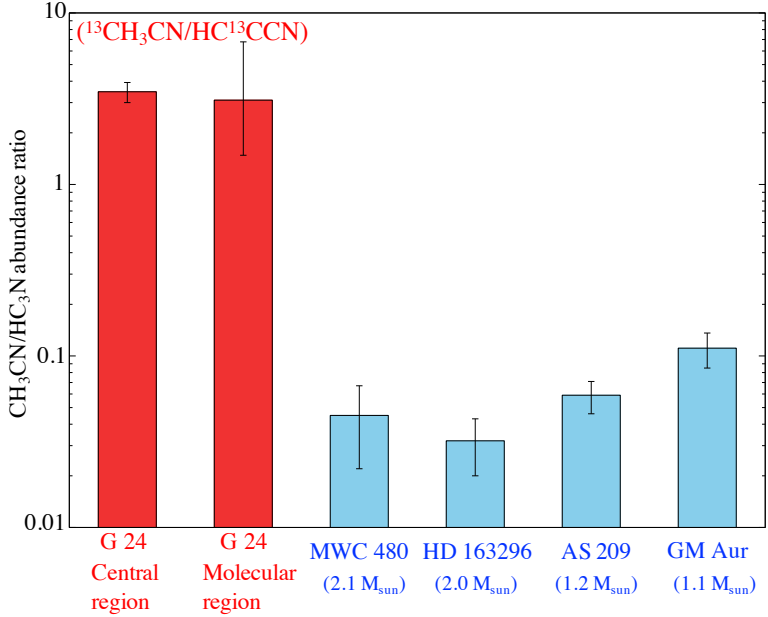}
 \end{center}
\caption{Comparisons of the CH$_{3}$CN/HC$_{3}$N abundance ratio among disks with different stellar masses. Values except for G24 are taken from \citet{2021ApJS..257....9I}. Stellar masses of Herbig Ae and T Tauri stars are taken from \citet{2021ApJS..257....1O}. \label{fig:ratio}} 
\end{figure*}

We compare the CH$_{3}$CN/HC$_{3}$N abundance ratios in the disk structure of G\,24 to those in disks around lower-mass stars, i.e., Herbig Ae stars and T Tauri stars, and investigate chemical properties of the molecular disk around G\,24.

In disks around lower-mass stars, a main formation mechanism of CH$_{3}$CN is considered to be dust-surface reactions \citep{2018ApJ...857...69B, 2018ApJ...859..131L}; (1) the successive hydrogenation reaction of C$_{2}$N and (2) a radical-radical reaction between CH$_{3}$ and CN.
Gas-phase routes have been also investigated \citep{2018ApJ...857...69B, 2018ApJ...859..131L}:
\begin{equation} \label{equ:gas1}
{\rm {CH}}_{3}^{+} + {\rm {HCN}} \rightarrow ({\rm {CH}}_{3}{\rm {NCH}}^{+})^{*} \rightarrow {\rm {CH}}_{3}{\rm {CNH}}^{+} + h\nu,
\end{equation}
followed by 
\begin{equation} \label{equ:gas2}
{\rm {CH}}_{3}{\rm {CNH}}^{+} + {\rm {e}}^{-} \rightarrow {\rm {CH}}_{3}{\rm {CN}} + {\rm {H}}.
\end{equation}
However, \citet{2018ApJ...859..131L} concluded that the dust-surface routes are more efficient than the gas-phase ones.
The CH$_{3}$CN formed on dust surfaces sublimates into the gas phase by thermal desorption or photo-evaporation.
As the excitation temperatures of CH$_{3}$CN are lower than its sublimation temperature in disks around lower-mass stars, the primary path would be photo-evaporation in such disks.

On the other hand, for HC$_{3}$N, only gas-phase formation routes are known, including the ion-molecule reactions and the neutral-neutral reactions \citep{2018ApJ...859..131L,2019ApJ...886...86L}.
For example, the following reactions are considered to contribute to the HC$_{3}$N formation:
\begin{equation} \label{equ:gas3}
{\rm {HC}}_{3}{\rm {NH}}^{+} + {\rm {e}}^{-} \rightarrow {\rm {HC}}_{3}{\rm {N}} + {\rm {H}},
\end{equation}
or
\begin{equation} \label{equ:gas4}
{\rm {C}}_{2}{\rm {H}}_{2} + {\rm {CN}} \rightarrow {\rm {HC}}_{3}{\rm {N}} + {\rm {H}}.
\end{equation}
The HC$_{3}$NH$^{+}$ ion is formed by various ion-molecule reactions \citep[e.g., C$_{2}$H$_{2}^{+}$+HCN, C$_{3}$H$_{n}^{+}$+N ($n=3,4,5$);][]{2016ApJ...830..106T,2017ApJ...846...46T}.

Both CH$_{3}$CN and HC$_{3}$N are destroyed by the UV photodissociation:
\begin{equation} \label{equ:gas5}
{\rm {CH}}_{3}{\rm {CN}} + h\nu \rightarrow {\rm {CH}}_{3} + {\rm {CN}},
\end{equation}
and 
\begin{equation} \label{equ:gas6}
{\rm {HC}}_{3}{\rm {N}} + h\nu \rightarrow {\rm {CCH}} + {\rm {CN}},
\end{equation}
respectively.
HC$_{3}$N is also destroyed by ions, such as H$^{+}$, H$_{3}^{+}$, HCO$^{+}$, in higher $A_{\rm {v}}$ regimes \citep[$A_{\rm {v}} \geq 4$ mag;][]{2019ApJ...886...86L}.
The reaction with C$^{+}$ could destroy CH$_{3}$CN \citep{2018ApJ...859..131L}, but its contribution is not dominant in the chemical network simulation conducted by \citet{2019ApJ...886...86L}.

Figure \ref{fig:ratio} shows comparisons of the CH$_{3}$CN/HC$_{3}$N abundance ratio among protostellar disks with different stellar masses. 
The abundance ratios in Herbig Ae and T Tauri stars are taken from \citet{2021ApJS..257....9I}.
MWC\,480 and HD\,163296 are Herbig Ae stars, and AS\,209 and GM Aur are T Tauri stars, respectively.
The derived $N$($^{13}$CH$_{3}$CN)/$N$(HC$^{13}$CCN) ratios around G\,24 ($\sim 3.0-3.5$) are higher than the CH$_{3}$CN/HC$_{3}$N abundance ratios in the other disks ($\sim0.03-0.11$) by more than one order of magnitude.
As discussed in Section \ref{sec:CH3CN/HC3N}, there are differences in the CH$_{3}$CN/HC$_{3}$N ratio around the G\,24 HC \ion{H}{2} region by a factor of a few among different methods, but the CH$_{3}$CN/HC$_{3}$N ratios around G\,24 are clearly higher than those in the disks around the lower-mass stars.

One possible explanation for the high CH$_{3}$CN/HC$_{3}$N abundance ratio around G\,24 is that the thermal sublimation mechanism enhances the gas-phase CH$_{3}$CN abundance around the massive star, while the photo-evaporation is important for sublimation of CH$_{3}$CN in the other disks.
In addition, HC$_{3}$N could be more efficiently destroyed around the massive star, because of the higher UV photodissociation rate of HC$_{3}$N than that of CH$_{3}$CN, as we have already discussed in Section \ref{sec:CH3CN/HC3N}.
HC$_{3}$N is also destroyed by reactions with ions (H$^{+}$, H$_{3}^{+}$, HCO$^{+}$), which should be abundant in ionized regions.

We also compare the CH$_{3}$CN/HC$_{3}$N ratios around G\,24 to that of the Orion Hot Core \citep[i.e., envelope scale;][]{2014ApJ...787..112C}.
The Orion Hot Core is associated with Source I, and its mass was estimated to be around 15 $M_{\sun}$ \citep{2020ApJ...889..178B}.
In addition, the disk structure has been detected around this source \citep{2020ApJ...889..155W}.
Thus, it seems to be a good source to compare the chemical composition to G\,24.
The CH$_{3}$CN and HC$_{3}$N abundances with respect to H$_{2}$ at the Orion Hot Core were derived to be $3.0 \times 10^{-8}$ and $8.1 \times 10^{-9}$, respectively \citep[$N$(H$_{2}$)=$3.1 \times 10^{23}$ cm$^{-2}$;][]{2014ApJ...787..112C}.
Hence, the CH$_{3}$CN/HC$_{3}$N abundance ratio is calculated at 3.7.
This value is comparable to the CH$_{3}$CN/HC$_{3}$N ratios around G\,24 ($\sim 3.0-3.5$).
The similar CH$_{3}$CN/HC$_{3}$N ratios between the Central and Molecular regions in G\,24, and between G\,24 and the Orion Hot Core imply the following two possibilities:
\begin{enumerate}
    \item The disk chemistry may be inherited from the envelope.
    \item The chemical processes at the envelope scale may be similar to those at the disk scale due to the powerful central source.
\end{enumerate}
Here, we demonstrate the possibility of the different nitrile chemistry between massive stars and lower-mass stars, i.e., Herbig Ae and T Tauri stars.
However, we have data of both CH$_{3}$CN and HC$_{3}$N toward only one massive star. 
In order to confirm the nitrile chemistry in disk structures around massive stars, we need to increase source samples.
Future observations covering several lines, which enable us to derive accurate column densities and excitation temperatures of CH$_{3}$CN and HC$_{3}$N, are necessary to understand the disk chemistry around massive stars.
These observations will help us to reveal the massive star formation processes.

\subsection{The possibility of a binary system} \label{sec:binary}

\begin{figure*}
\begin{center}
\includegraphics[width=\textwidth]{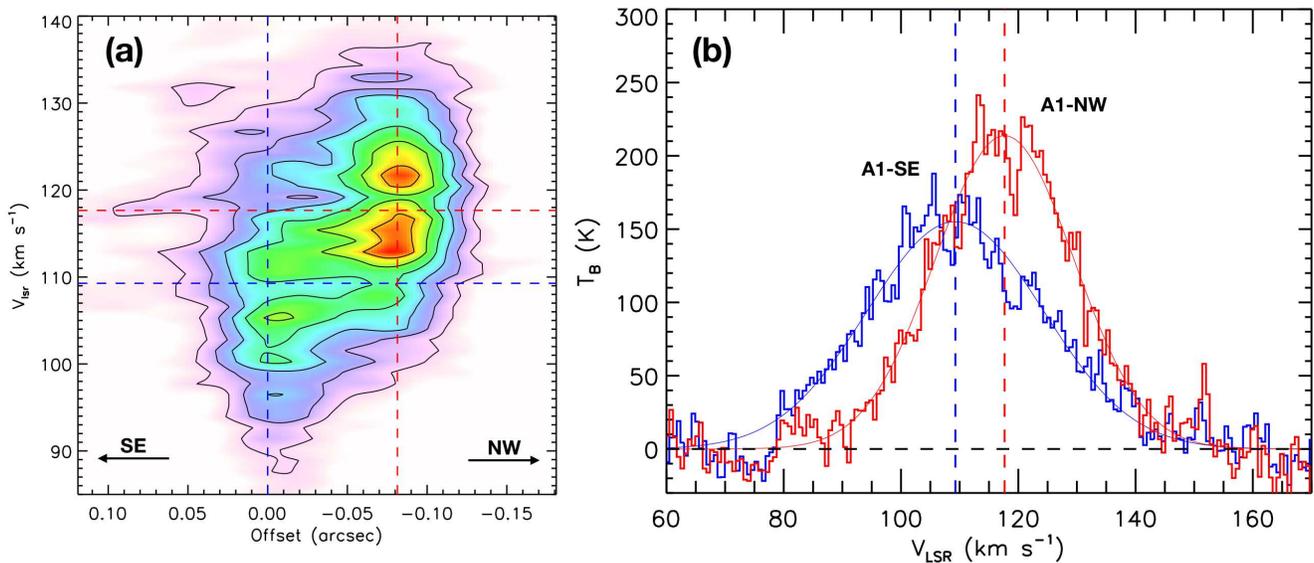}\\
\caption{{\bf (a):} 
Position-velocity diagram of the H30$\alpha$ emission along a strip connecting the
two continuum peaks with a width of 0.01$\arcsec$.
The dashed lines in red and blue colors mark the locations and center velocities
of the southeast peak and the northwest peak, respectively.
{\bf (b):}
H30$\alpha$ spectra at the two continuum peak positions 
(blue: southeast peak, named as A1-SE, red: northwest peak, named as A1-NW).
The solid thin lines show the Gaussian fitting to the spectra,
with the central velocities marked by the vertical dashed lines
($v_\mathrm{center}=109.3\pm 0.3$\,km\,s$^{-1}$ for A1-SE 
and $v_\mathrm{center}=117.7\pm 0.2$\,km\,s$^{-1}$ for A1-NW).
}
\label{fig:binary}
\end{center}
\end{figure*}

As shown in panel (a) of Figure \ref{fig:mom0}, the two peaks with the separation of $\sim0.1\arcsec$ are detected in the 1.38 mm continuum emission.
Panel (a) of Figure \ref{fig:binary} shows the P-V diagram of the H30$\alpha$ emission along
a direction connecting the two continuum peaks ($\mathrm{PA}=-31^\circ$). 
It clearly shows that the H30$\alpha$ emission associated with the southeast peak is blueshifted by $\sim10{\rm\:km\:s^{-1}}$ from the northwest peak.

\citet{2021A&A...650A.142M} interpreted this velocity gradient as the ionized disk rotation around a massive protostar.
Assuming an edge-on orientation, they derived the dynamical mass of $\sim17~M_\odot$, which should be the minimum mass considering the possible inclination.
Consistent with this dynamical mass evaluation,
\citet{2019A&A...624A.100C} suggested an O9.5 star based on the derived Lyman continuum rate of $5.3 \times 10^{47}{\rm\:s}^{-1}$.
We also constructed the spectral energy distribution (SED) toward this source to retrieve the stellar mass and other physical properties,
using infrared data from Spitzer, WISE, and Herschel (Table \ref{tab:SED_flux_densities} in Appendix \ref{sec:a3}) and fitting the SED with the \citet{2018ApJ...853...18Z} model
(see Appendix \ref{sec:a3} for the details).
The best models of the SED fitting suggested the stellar masses of $16$--$24\:M_\odot$ with the bolometric luminosities of a few times $10^5\:L_\odot$ (see Figure \ref{fig:SED} and Table \ref{tab:SED} in Appendix \ref{sec:a3}),
which is also consistent with the dynamical mass estimation with the Keplerian disk by \citet{2021A&A...650A.142M}.
The agreement of the independent analyses supports the presence of a $\sim20\:M_\odot$ star with the ionized-gas disk in the center of the G\,24 HC \ion{H}{2} region.

However, this ionized disk scenario does not explain the presence of the two continuum peaks well.
An alternative interpretation would be that the kinematic structure of H30$\alpha$ consists of two ionized disks associated with two massive young stars orbiting around each other, i.e., a binary system.
In fact, instead of one single velocity gradient across $\sim 0.1\arcsec$, the P-V diagram appears to be more consistent with two separate components with the velocity offset (i.e., the orbital motion), each of which has its own velocity gradient (i.e., the rotation of each disk).
It is worth briefly discussing the possibility of the binary system in G\,24, considering its asymmetric structures and the high binary rate of massive stars.

We can derive the total mass of the binary system assuming that the protostars locate at the continuum peaks and their velocity offset is due to the orbital motion.
The peak separation is $0.08\arcsec$ or $\Delta S\simeq540{\rm\:au}$, which is a project separation.
We estimate their velocity offset as $\Delta V\simeq 8.4 {\rm\:km\:s^{-1}}$
from the H30$\alpha$ spectra at the peak locations (panel (b) of Figure \ref{fig:binary}).
Considering a circular Kepler orbit, we derive the total dynamical mass as $M_{\rm tot}=\Delta V^2 \Delta S / G \simeq 43\:M_\odot$, where $G$ is the gravitational constant.
This formula is different from that in the disk scenario by a factor of 8, i.e., $M_{\rm c}=\Delta V^2 \Delta S /(8G)$ \citep[where $M_c$ is the stellar mass at the disk center; see Sec. 4.1 of][]{2021A&A...650A.142M},
because $\Delta S$ represents the separation of the two sources while it is the diameter of the ionized disk in the disk scenario.
%The actual positions of the binary stars could have offset from the continuum peaks, because free-free and H30$\alpha$ emissions extend from the peak positions.
Since free-free and H30$\alpha$ emissions trace the surrounding ionized gas rather than the protostar itself, the actual positions of the protostars might be slightly off the continuum emission peaks.
Considering such $\Delta S$ uncertainty of $\sim100{\rm\:au}$ (or $0.015\arcsec$) and the $\Delta V$ fitting error of $0.36{\rm\:km\:s^{-1}}$,
we estimate that the mass range of $M_{\rm tot}\simeq34$--$51\:M_\odot$ is consistent with the binary scenario.
We note that the circular orbit perpendicular to the plane of sky are assumed here, and thus we consider those as the minimum masses.

The total values of the Lyman continuum rate and the infrared luminosity can also provide constraints on the stellar masses of the binary.
\citet{2018A&A...616A..66M} and \citet{2019A&A...624A.100C} estimated a total Lyman continuum rate of $(5.3$--$7.2)\times 10^{47}~\mathrm{s}^{-1}$ by fitting the radio to millimeter SED.
Since the two peaks have similar brightness in continuum and H30$\alpha$ total emissions,
the two massive stars in the binary should have similar masses, and therefore similar Lyman continuum rate of $\sim3\times 10^{47}~\mathrm{s}^{-1}$.
Such a Lyman continuum rate corresponds to a zero-age main sequence (ZAMS) mass of $18~M_\odot$, making a total mass of $36~M_\odot$ for the binary, which is close to the minimum dynamical mass estimated above.
If the binary stars have not reached the ZAMS phase, their masses can be higher because their Lyman continuum rate would be lower than those in the ZAMS phase at the same masses \citep[e.g.,][]{2016ApJ...818...52T}.
As discussed in Appendix \ref{sec:a3}, we conducted the infrared SED fitting.
The mass estimation from this fitting is not applicable for the binary system because it assumed a single-star system.
However, the obtained total bolometric luminosity of a few $\times 10^5~L_\odot$ is reasonable even for the multiple-source system (Table \ref{tab:SED}).
Considering the protostellar evolution, the luminosity of a massive protostar with $\sim15$--$20~M_\odot$ would be as bright as $10^5~L_\odot$ \citep[e.g.,][]{2018ApJ...853...18Z},
which is again consistent with the estimation of the dynamical mass.

The systemic velocity of G\,24 measured from the outer molecular emissions is $\sim$112\,km\,s$^{-1}$.
If we assume that the line-of-sight velocity of the mass center of the binary is $112{\rm\:km\:s^{-1}}$,
the mass ratio between the primary and secondary would be about 2:1
from the central velocity of each continuum peak.
Considering the dynamical mass of $>37~M_\odot$,
we evaluate $\gtrsim24~M_\odot$ for A1-SE and $\gtrsim12~M_\odot$ for A1-NW.
However, such analysis is susceptible to the determination of the line center velocities of H30$\alpha$, which have FWHMs of $30-35$\,km\,s$^{-1}$, as well as the binary system velocity.
Therefore, it is still difficult to accurately determine the masses of the individual stars via dynamics.

We note that the binary scenario does not necessarily contradict the other features of G\,24.
In the case of the binary star scenario, the rotating structure traced by CH$_3$CN and HC$_3$N is a circumbinary disk surrounding the two massive stars.
The P-V diagrams of the molecular lines are roughly consistent with the Keplerian rotation with $\sim20\:M_\odot$,
but the disk is assumed to be perpendicular to the sky in these diagrams
\citep[Figure \ref{fig:PV}; Figure 4 of][]{2021A&A...650A.142M}.
Hence, the value of $20\:M_\odot$ is the minimum enclosed mass, and the total dynamical mass of $>37~M_\odot$ from the H30$\alpha$ emission is acceptable.
The presence of a single jet perpendicular to the molecular disk reported by \citet{2018A&A...616A..66M, 2021A&A...650A.142M} does not exclude the binary scenario.
For example, in the forming massive binary system IRAS\,16547--4247,
one of the binary stars launches a single jet nearly perpendicular to the circumbinary disk,
while the other star shows no jet feature, despite both protostars show almost the same levels of continuum and line emissions \citep{2020ApJ...900L...2T}.
If this is also happening in G\,24, one star may be supplied with more material from the circumstellar disk.
Such an asymmetric infalling structure at $<1000{\rm\:au}$ could cause the asymmetric UV field suggested in the larger scale.

Similar to the binary scenario for G\,24,
\citet{2019NatAs...3..517Z} reported a forming massive binary IRAS\,07299--1651
with each member traced by hydrogen recombination lines showing velocity offset caused by orbital motion.
In IRAS 07299-1651, the line emissions from the two stars are well separated that allows analyzing the circumstellar disk rotation of the primary star.
However, in G\,24, the H30$\alpha$ emissions from the continuum peaks have already spread out and merged together,
making it difficult to analyze the detailed kinematic structures of the ionized gas,
i.e., the circumstellar disks and their circumbinary disk.
We speculate that G\,24 is more massive than IRAS 07299-1651, and thus the ionized gas is more extended.
Therefore, although we suggest the possibility of the binary system in G\,24,  we cannot confirm such a scenario with the current data set.
Future higher-angular-resolution observations may help to confirm this point.

\section{Conclusions} \label{sec:con}

We have analyzed the ALMA archival data in Band 6 toward the G\,24 HC \ion{H}{2} region.
The vibrationally-excited lines of HC$_{3}$N ($v_{7}=2$, $J=24-23$) have been detected around the G\,24 HC \ion{H}{2} region.
Main results and conclusions of this paper are as follows.

\begin{enumerate}
\item We have compared the moment 0 map and the P-V diagram of the HC$_{3}$N ($v_{7}=2$, $J=24-23$, $l=2e$) line and the CH$_{3}$CN ($v_{8}=1$, $J_{K,l}=12_{6,1}-11_{6,1}$) line. 
Features in the spatial distributions and the P-V diagram of HC$_{3}$N are similar to those of CH$_{3}$CN.
Thus, the HC$_{3}$N emission is tracing the molecular disk around the G\,24 HC \ion{H}{2} region previously identified by the CH$_{3}$CN lines.
These results indicate that the HC$_{3}$N emission can be used as a disk tracer for massive protostars even at the evolutionary stage of the HC \ion{H}{2} region.

\item We have derived the column density ratios of $^{13}$CH$_{3}$CN/HC$^{13}$CCN at the two representative regions, the Central and Molecular regions, which are determined based on the continuum map and the HC$_{3}$N moment 0 map, respectively.
The ratios are derived to be $3.47 \pm 0.46$ and $3.10_{-1.62}^{+3.68}$ at the Central and Molecular regions, respectively.
We have also derived the integrated-intensity ratios between the CH$_{3}$CN ($v_{8}=1$, $J_{K,l}=12_{6,1}-11_{6,1}$) and HC$_{3}$N ($v_{7}=2$, $J=24-23$, $l=2e$) lines and between the $^{13}$CH$_{3}$CN ($J_{K}=13_{3}-12_{3}$) and HC$^{13}$CCN ($J=26-25$) lines.
All of the CH$_{3}$CN/HC$_{3}$N ratios derived based on integrated intensity and column density show the higher values at the Central region than those at the Molecular region.
These results suggest that HC$_{3}$N is more efficiently destroyed in the region irradiated by the strong UV radiation.

\item We have compared the $^{13}$CH$_{3}$CN/HC$^{13}$CCN abundance ratios around G\,24 to the CH$_{3}$CN/HC$_{3}$N abundance ratios in the disks around Herbig Ae and T Tauri stars.
The abundance ratios around G\,24 are higher than those in the other disks by more than one order of magnitude. 
The difference between the G\,24 HC \ion{H}{2} region and the other disks is caused by (1) efficient thermal desorption of CH$_{3}$CN in hot and dense region around the G\,24 HC \ion{H}{2} region, and (2) rapid destruction of HC$_{3}$N in the region irradiated by the strong UV radiation around the G\,24 HC \ion{H}{2} region.

\item Based on the two peaks of the free-free emission and their H30$\alpha$ kinematics in the central ionized region,
we briefly discussed the possibility that it is composed of a binary system.
We evaluated the total dynamical mass is $\gtrsim37~M_\odot$,
which is consistent with the total bolometric luminosity and the Lyman continuum.
However, we cannot confirm the binary scenario with the current data set.
This is because the H30$\alpha$ emissions extended from the two continuum peaks merge in the Position-Velocity space, which makes the detailed dynamical analysis challenging.

\end{enumerate}

We have shown that HC$_{3}$N can be used as a disk tracer even for massive protostars.
The nitrile species, CH$_{3}$CN and HC$_{3}$N, are mainly formed by different formation processes (dust-surface reactions vs. gas-phase reactions), and thus, the CH$_{3}$CN/HC$_{3}$N abundance ratio will be able to become a useful tracer for physical structures of disks around massive protostars.
We need to increase source samples of massive stars as well as lower-mass stars, in order to understand the nitrile chemistry and the connection between physics and chemistry in disk structures.

\begin{acknowledgments}
This paper makes use of the following ALMA data: ADS/JAO.ALMA\#2018.1.00745.S.
ALMA is a partnership of ESO (representing its member states), NSF (USA) and NINS (Japan), together with NRC (Canada), MOST and ASIAA (Taiwan), and KASI (Republic of Korea), in cooperation with the Republic of Chile.
The Joint ALMA Observatory is operated by ESO, AUI/NRAO and NAOJ.
The National Radio Astronomy Observatory is a facility of the National Science Foundation operated under cooperative agreement by Associated Universities, Inc.
Data analysis was in part carried out on the Multi-wavelength Data Analysis System operated by the Astronomy Data Center (ADC), National Astronomical Observatory of Japan.
We thank the anonymous referee whose valuable comments helped improve the paper.  

K.T. appreciates the support by JSPS KAKENHI Grant No. JP20K14523. 
K.T. thanks Dr. Yusuke Miyamoto (East Asian ALMA Regional Center/National Astronomical Observatory of Japan) for his kind support for our data reduction at the ADC.
K.E.I.T. acknowledges the support by JSPS KAKENHI Grant Numbers JP19K14760, JP19H05080, JP21H00058, and JP21H01145.
R.F. acknowledges funding from the European Union's Horizon 2020 research and innovation program under the Marie Sklodowska-Curie grant agreement No 101032092.
JCT acknowledges support from ERC project MSTAR and VR grant 2017-04522.
S.T. is supported by JSPS KAKENHI Grant Numbers JP21H00048 and JP21H04495. 
L.M. acknowledges the financial support of DAE and DST-SERB research grants (SRG/2021/002116 and MTR/2021/000864) of the Government of India.
\end{acknowledgments}


\vspace{5mm}
\facilities{Atacama Large Millimeter/submillimeter Array (ALMA)}

\software{Common Astronomy Software Applications package \citep[CASA;][]{2007ASPC..376..127M}, 
Astropy \citep{astropy2013, astropy2018}; this research made use of Photutils, version 1.0.0, an Astropy package for detection and photometry of astronomical sources \citep{2020zndo...4044744B}.}



\appendix

\section{Moment 0 maps of the $^{13}$C isotopologues} \label{sec:a1}

\begin{figure*}
 \begin{center}
  \includegraphics[bb =0 25 770 545, scale=0.65]{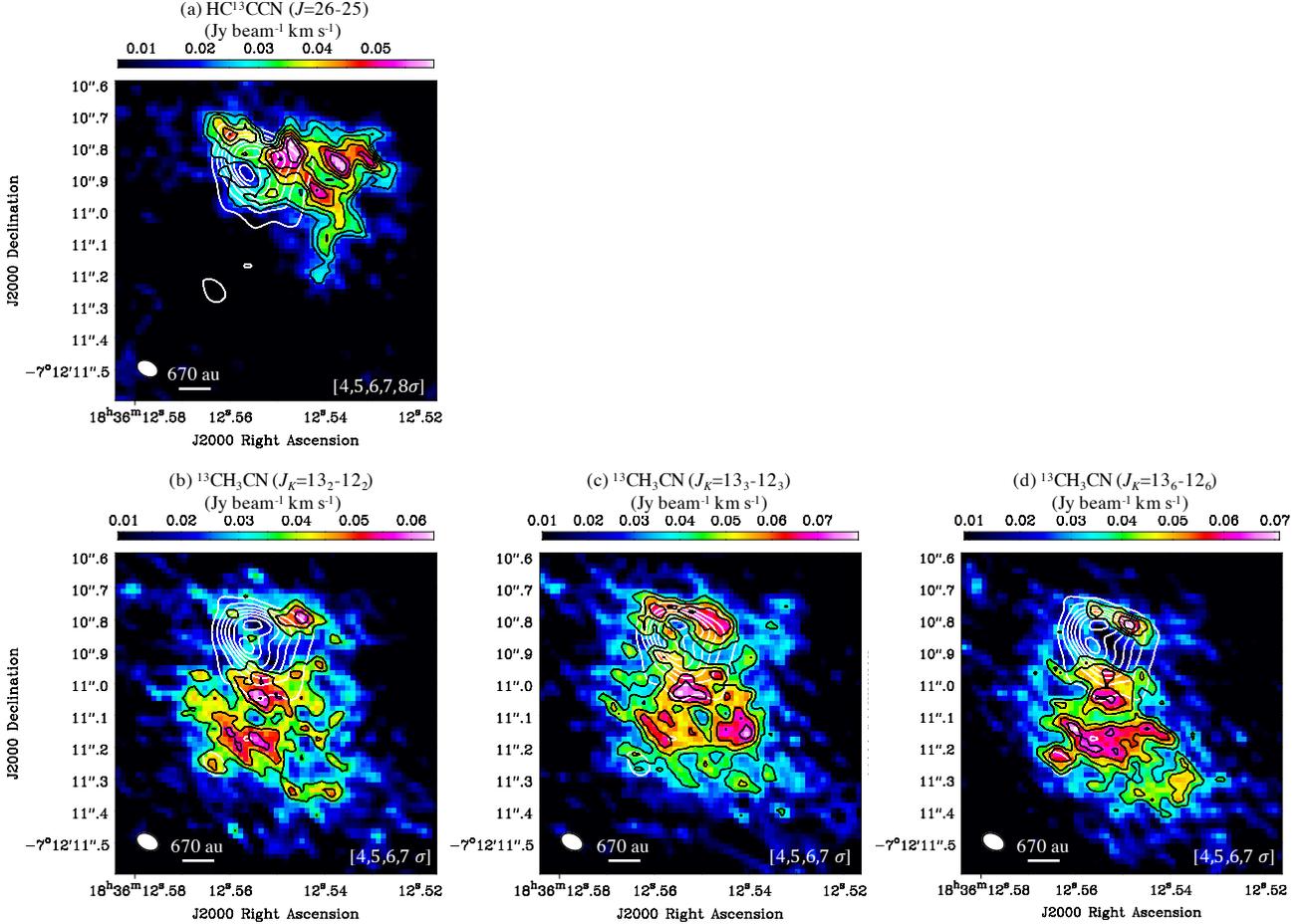}
 \end{center}
\caption{Moment 0 images of (a) HC$^{13}$CCN ($J=26-25$), (b)$^{13}$CH$_{3}$CN ($J_{K}=13_{2}-12_{2}$), (c)$^{13}$CH$_{3}$CN ($J_{K}=13_{3}-12_{3}$), and (d)$^{13}$CH$_{3}$CN ($J_{K}=13_{6}-12_{6}$), respectively. The rms noise levels are 7 mJy\,beam$^{-1}$ km\,s$^{-1}$ for panel (a), 9 mJy\,beam$^{-1}$ km\,s$^{-1}$ for panels (b) and (d), and 10 mJy\,beam$^{-1}$ km\,s$^{-1}$ for panel (c), respectively. The black contour levels are indicated in each panel. The white contours indicate the continuum image from 20$\sigma$ to 80$\sigma$, in steps of 10$\sigma$. The filled white ellipses indicate angular resolution of 0\farcs068 $\times$ 0\farcs048 for panel (a), and 0\farcs074 $\times$ 0\farcs051 for the other panels, respectively. The linear scale is given for 0\farcs1 corresponding to 670 au.\label{fig:13C_mom0}}
\end{figure*}

Figure \ref{fig:13C_mom0} shows moment 0 maps of HC$^{13}$CCN and $^{13}$CH$_{3}$CN toward the G\,24 HC \ion{H}{2} region.
We only show three lines of $^{13}$CH$_{3}$CN ($J_{K}=13_{K}-12_{K}$, $K=2,3,6$), because these lines are not contaminated by other lines.

\section{Channel maps of the vibrationally-excited lines} \label{sec:a2}

\begin{figure*}
 \begin{center}
 \includegraphics[bb =20 10 481 592, scale=1.0]{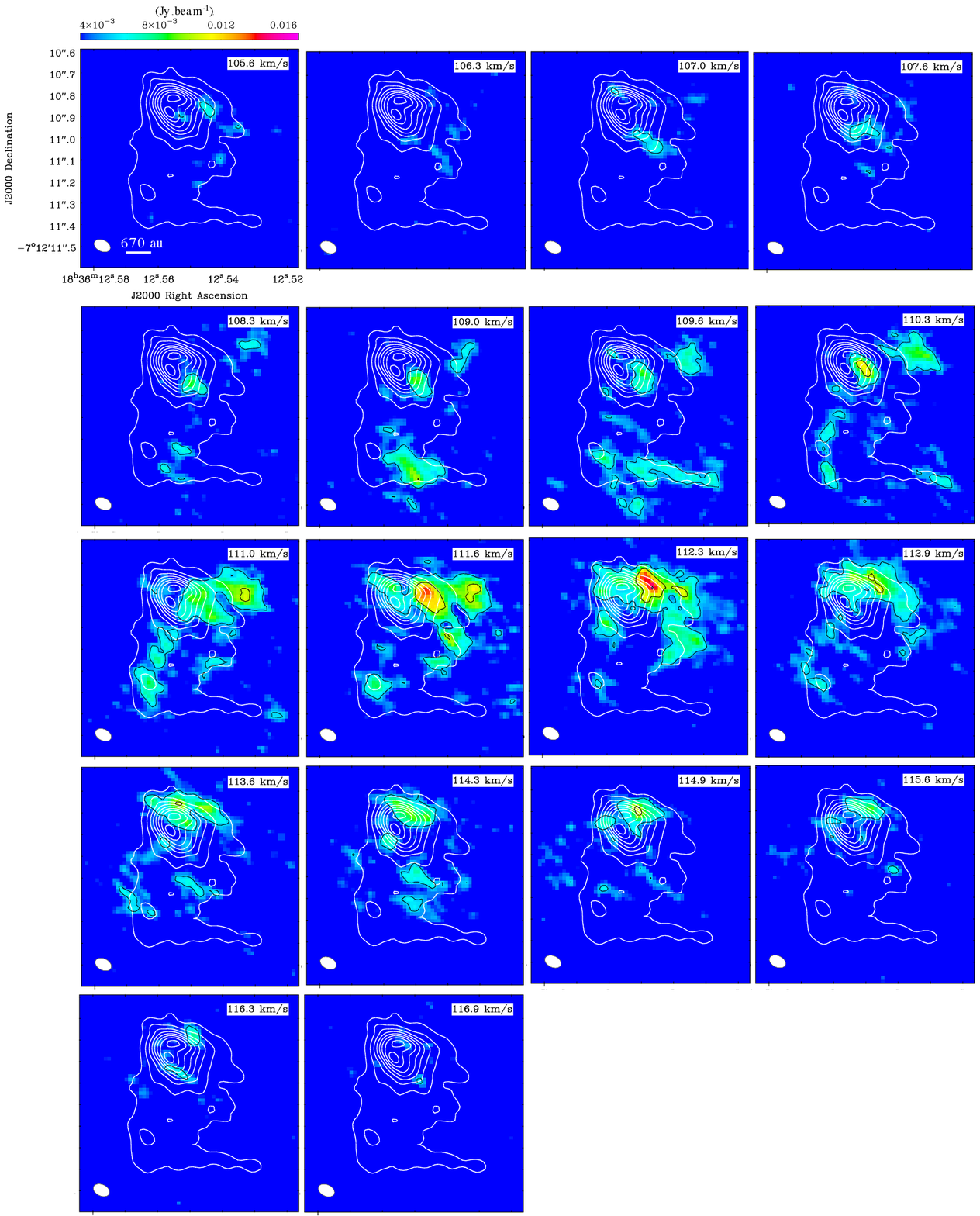}
 \end{center}
\caption{Channel maps of the HC$_{3}$N ($v_{7}=2$, $J=24-23$, $l=2e$) line ranging from 105.6 km\,s$^{-1}$ to 116.9 km\,s$^{-1}$. The black contours indicate the signal levels from $5\sigma$ in steps of $5\sigma$. The noise level is 1.1 mJy\,beam$^{-1}$. The white contours indicate the continuum image from 10$\sigma$ to 80$\sigma$, in steps of 10$\sigma$.  The filled white ellipses indicate angular resolution of 0\farcs079 $\times$ 0\farcs053. The linear scale is given for 0\farcs1 corresponding to 670 au.\label{fig:channelHC3N}}
\end{figure*}

\begin{figure*}
 \begin{center}
  \includegraphics[scale=1.0]{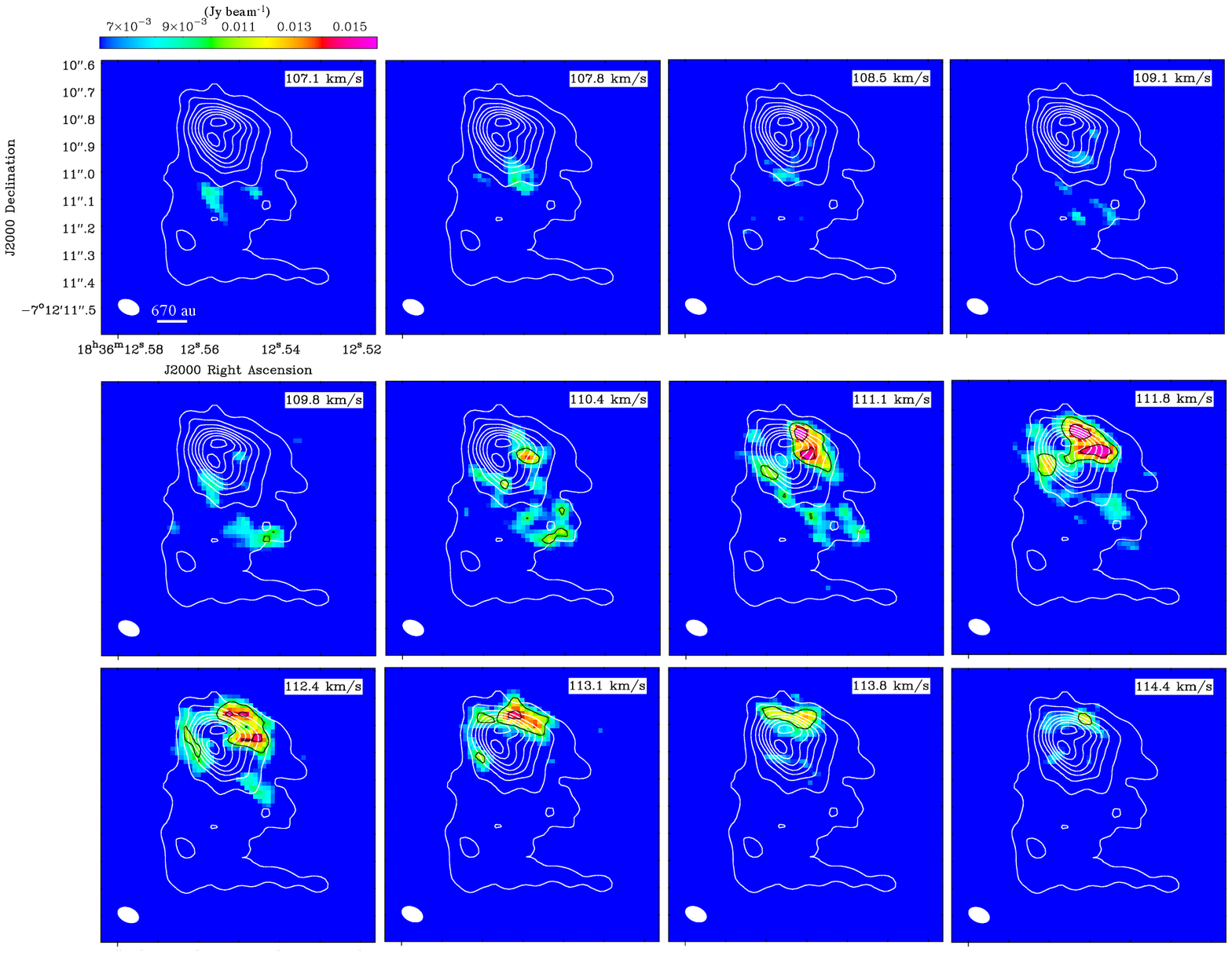}
 \end{center}
\caption{Channel maps of the CH$_{3}$CN ($v_{8}=1$, $J_{K,l}=12_{6,1}-11_{6,1}$) line ranging from 107.1 km\,s$^{-1}$ to 115.1 km\,s$^{-1}$. The black contours indicate the signal levels from $5\sigma$ in steps of $2\sigma$. The noise level is 2.0 mJy\,beam$^{-1}$. The white contours indicate the continuum image from 10$\sigma$ to 80$\sigma$, in steps of 10$\sigma$.  The filled white ellipses indicate angular resolution of 0\farcs082 $\times$ 0\farcs053. The linear scale is given for 0\farcs1 corresponding to 670 au.\label{fig:channelCH3CN}}
\end{figure*}

Figures \ref{fig:channelHC3N} and \ref{fig:channelCH3CN} show channel maps of the vibrationally-excited lines of HC$_{3}$N ($v_{7}=2$, $J=24-23$, $l=2e$) and CH$_{3}$CN ($v_{8}=1$, $J_{K,l}=12_{6,1}-11_{6,1}$), respectively.

\section{SED fitting toward the G\,24 HC \ion{H}{2} region} \label{sec:a3}

In order to retrieve physical information from the HC \ion{H}{2} region, we constructed the spectral energy distribution (SED) and subsequently fitted the \citet{2018ApJ...853...18Z} model grid. 
We first measured the flux density in Spitzer, WISE, and Herschel data (Table\,\ref{tab:SED_flux_densities}). 
To do this, we performed circular aperture photometry fixing the aperture radius to 16\arcsec\ to all wavelengths, following the fiducial method in \citet{2017ApJ...843...33D,2019ApJ...874...16L,2020ApJ...904...75L}. 
The aperture size was chosen to enclose most of the flux in the 70\,\micron\ Herschel image and variations of 30\% in the aperture radius to larger size did not affect significantly the final flux obtained. 
The standard method in performing circular aperture photometry also subtracts the background emission that is evaluated in an annulus from one to two aperture radii. 
%with  inner radius equal to the aperture radius and outer radius equal to two times the aperture radius. 
The background-subtracted flux densities are reported in the second column of Table\,\ref{tab:SED_flux_densities} and are the ones used in the SED fitting (we also report in the same column the flux densities without background subtraction in parenthesis). 
Once the fluxes were measured, we fitted to the SED model grid that provides estimates of key protostellar properties. We used the \citet{2018ApJ...853...18Z} model grid and developed an improved version written in python called \textit{sedcreator}\footnote{\url{https://github.com/fedriani/sedcreator}\\or \url{https://pypi.org/project/sedcreator/}} based on the original code written in IDL\footnote{\url{https://zenodo.org/record/1134877\#.YRJl4ZMza84}}. 
The details of this package will be publicly available in a forthcoming paper (Fedriani et al. in prep.). 
The \citet{2018ApJ...853...18Z} model grid is based on the assumption that massive stars are formed from massive prestellar cores supported by internal pressure. 
The model grid self-consistently includes the evolutionary sequences of protostar, disk, envelope, and outflow cavity. 
The constrained free parameters of the model grid are: initial core mass ($M_\mathrm{c}$), environmental clump mass surface density ($\Sigma_\mathrm{cl}$), current protostellar mass ($m_*$), viewing angle with respect to the outflow axis ($\theta_\mathrm{view}$) and amount of foreground extinction ($A_\mathrm{V}$).

\begin{figure*}[!th]
 \begin{center}
 \includegraphics[bb =20 20 705 462, scale=0.6]{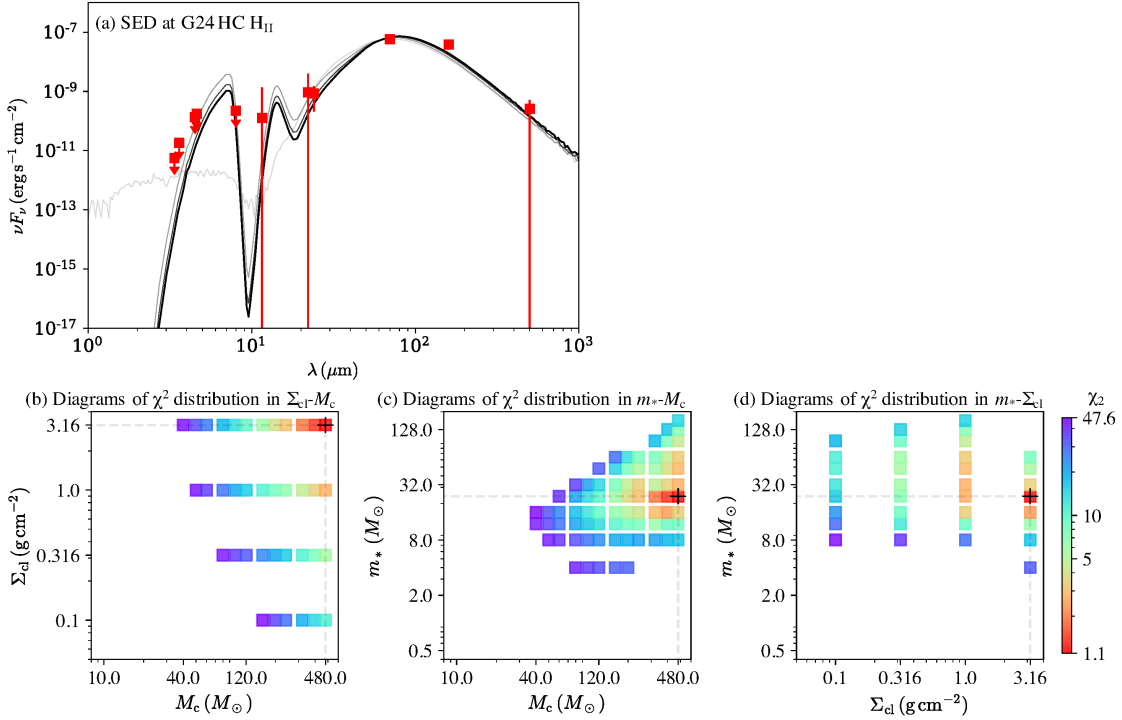}
 \end{center}
\caption{Panel (a) SEDs of the 5-best models with the observations represented as red squares. The best-fit model is shown with a solid black line and the next four best models are shown with solid gray lines. Panels (b) -- (d) Diagrams of $\chi^{2}$ distribution in $\Sigma_{\rm cl}$-$M_{\rm {c}}$ space, $m_{*}$-$M_{\rm{c}}$ space, and $m_{*}$-$\Sigma_{\rm cl}$ space, respectively. The black crosses indicate the locations of the best model. \label{fig:SED}}
\end{figure*}

By minimizing the $\chi^2$ function defined in Eq. 4 of \citet{2018ApJ...853...18Z} for each physical model, we find that the best five models are consistent with a $M_\mathrm{c}$ ranging from 320 to 480\,$M_\odot$, a $\Sigma_\mathrm{cl}$ with 1.0-3.16\,$\mathrm{g\,cm^{-2}}$, and, most importantly, a current protostellar mass of $16-24$\,$M_\odot$ (Table\,\ref{tab:SED}).
The intrinsic bolometric luminosity is also a few $\times 10^5$\,L$_\odot$, i.e., which is the level associated with such massive protostars.
%hinting the high mass of the central protostar. 
This further supports the findings from the P-V analysis from the molecular emission. Figure\,\ref{fig:SED} panel a shows the best five models, with the best model represented with thick black line, and the observations as red squares. 
The error bars are set to be the larger of either 10\% of the clump background-subtracted flux density to account for calibration error, or the value of the estimated clump background flux density. 
Note that all data points below $\lambda < 8\,\micron$ are treated as upper limits \citep[see, e.g.,][]{2017ApJ...843...33D}. 
Panels (b) and (c) of Figure\,\ref{fig:SED} show the 2D distribution of the three main physical parameters, i.e., $M_{\rm {c}}$, $\Sigma_{\rm cl}$, and $m_{*}$.

\begin{deluxetable}{lc}
\tablecaption{Background subtracted flux densities derived at each wavelength for the SED plot \label{tab:SED_flux_densities}}
\tablewidth{0pt}
\tablehead{
\colhead{Wavelength} & \colhead{Flux density} \\
\colhead{(\micron)} & \colhead{(Jy)}
}
\startdata
    \textit{Spitzer} & \\
    3.6 & 0.02 (0.13)\\
    4.5 & 0.20  (0.33)\\
    8.0 & 0.58  (4.13)\\
    24.0 & 6.81  (12.00)\\
    \noalign{\smallskip}
    \hline
    \noalign{\smallskip}
    \textit{WISE} & \\
    3.4 & 0.01   (0.08)\\
    4.6 & 0.27   (0.43)\\
    11.5 & 0.49   (5.31)\\
    22.0 & 6.92   (30.06)\\
    \noalign{\smallskip}
    \hline
    \noalign{\smallskip}
    \textit{Herschel} & \\
    70.0  & 1352.47   (1539.58)\\
    160.0 & 2046.43   (2487.86)\\
    500.0 & 43.32 (86.73)\\
\enddata
\tablecomments{Values in parenthesis are fluxes without the continuum subtraction. The photometric aperture radius was fixed at 16\arcsec, which was based on the $70\,\mu$m Herschel image.}
\end{deluxetable}

\begin{deluxetable*}{cccccccccccc}
\tablecaption{Parameters of the 5 best-fitted models \label{tab:SED}}
\tablewidth{0pt}
\tablehead{
\colhead{$\chi^2$} & \colhead{$M_\mathrm{c}$} & \colhead{$\Sigma_{\rm cl}$} & \colhead{$R_\mathrm{c}$} & \colhead{$m_{*}$} & \colhead{$\theta_\mathrm{view}$} & \colhead{$A_V$} & \colhead{$M_\mathrm{env}$} & \colhead{$\theta_\mathrm{w,esc}$} & \colhead{$\dot{M}_\mathrm{disk}$} & \colhead{$L_\mathrm{bol,iso}$} & \colhead{$L_\mathrm{bol}$} \\
\colhead{$\mathrm{}$} & \colhead{(${M_{\odot}}$)} & \colhead{($\mathrm{g\,cm^{-2}}$)} & \colhead{($\mathrm{pc}$)} & \colhead{(${M_{\odot}}$)} & \colhead{($\mathrm{{}^{\circ}}$)} & \colhead{($\mathrm{mag}$)} & \colhead{(${M_{\odot}}$)} & \colhead{($\mathrm{{}^{\circ}}$)} & \colhead{(${M_{\odot}\,yr^{-1}}$)} & \colhead{($\mathrm{L_{\odot}}$)} & \colhead{($\mathrm{L_{\odot}}$)}
}
%%\decimalcolnumbers
\startdata
1.11 & 480 & 3.160 & 0.09 & 24 & 13 & 361.92 & 440.54 & 12 & 2.0$\times10^{-3}$ & 1.1$\times10^6$ & 2.9$\times10^5$ \\
1.38 & 400 & 3.160 & 0.08 & 24 & 13 & 350.15 & 361.65 & 13 & 1.9$\times10^{-3}$ & 1.3$\times10^6$ & 3.0$\times10^5$ \\
1.93 & 320 & 3.160 & 0.07 & 24 & 13 & 325.19 & 276.82 & 15 & 1.8$\times10^{-3}$ & 1.8$\times10^6$ & 3.1$\times10^5$ \\
2.37 & 400 & 3.160 & 0.08 & 16 & 62 & 0.00   & 368.90 & 10 & 1.5$\times10^{-3}$ & 8.8$\times10^4$ & 1.0$\times10^5$ \\
2.72 & 480 & 1.000 & 0.16 & 24 & 22 & 217.99 & 433.43 & 15 & 8.2$\times10^{-4}$ & 1.9$\times10^5$ & 2.1$\times10^5$
\enddata
\end{deluxetable*}



\end{document}